\documentclass{aa}  

\usepackage{graphicx}
\usepackage{txfonts}

\usepackage[english]{babel}
\usepackage{subfigure}
\usepackage{amsmath} 
\usepackage{natbib}
\usepackage{xspace}

\usepackage[]{hyperref}
\begin{document}

\title{nDspec: A new Python library for modelling multi-dimensional datasets in X-ray astronomy}

\author{Matteo Lucchini \inst{1}
      \and
      Benjamin Ricketts \inst{1,2}
      \and 
      Phil Uttley \inst{1} 
      \and 
      Daniela Huppenkothen \inst{1}}

\institute{Anton Pannekoek Institute for Astronomy, University of Amsterdam,
           Science Park 904, 1098XH Amsterdam, the Netherlands
           \and
           SRON, Niel Bohrweg 4, 2033 CA Leiden, the Netherlands }

\abstract
{The current fleet of X-ray telescopes produces a wealth of multi-dimensional data, allowing us to study sources in time, photon energy, and polarisation. At the same time, it has become increasingly clear that progress in our physical understanding will only come from studying these sources in multiple dimensions simultaneously.} 
{Enabling multi-dimensional studies of X-ray sources requires new theoretical models predicting these datasets, new methods to analyse them, and, crucially, a software framework to combine data, models, and methods efficiently. However, the current ecosystem of software packages developed for X-ray data analysis does not provide the flexibility for advanced modelling of multi-dimensional datasets.}
{In this paper, we introduce \textsc{nDspec}, a new python-based library designed to allow users to seamlessly model one- and multi-dimensional datasets common to X-ray astronomy. Unlike most other libraries, it is designed as a flexible, modular, and extensible framework capable of accommodating multi-dimensional data and able to connect to a range of different inference tools and algorithms.}
{Here we focus on modelling timing and spectral-timing data as a function of both Fourier frequency and energy, in addition to limited support for time-averaged spectra. We discuss design philosophy and current features, and showcase an example use case by characterising a NICER observation of a black hole X-ray binary. We also highlight plans for extensions to other dimensions and new features, such as the inclusion of polarimetry and the improved statistical methods for Bayesian inference.}
{}


\maketitle
%
\section{Introduction}

The data collected from modern X-ray observatories is inherently multi-dimensional: X-ray detectors record energies, times of arrival, polarisation of incoming photons, and potentially the position on the sky. Thanks to the increased sensitivity of these observatories, it is possible for X-ray astronomers to study two or more of these quantities at the same time - for example spectral imaging (\citealt{deVries18}, \citealt{Vink22}), spectral-timing (\citealt{Uttley11,deMarco17,Wang22}), or soon polarimetric-timing \citep{Ingram17} studies. This kind of multi-dimensional analysis generally provides more information than traditional spectroscopy alone, and it can be invaluable for breaking modelling degeneracies and constraining the fundamental physics of a given system (e.g. \citealt{Mastroserio18}, \citealt{Ingram22Review}, \citealt{sushi}). 

Until recently, both datasets and models were simple enough to accommodate for one-dimensional modelling, even when considering multiple dimensions: for example, time-resolved spectroscopy relies on taking spectra at different points in time and modelling each individually (e.g. \citealt{Mastroserio18}, \citealt{Rogantini22}) with existing spectral modelling software (e.g. \textsc{Xspec} \cite{Arnaud96}, \textsc{ISIS} \citealt{Houck00}, \textsc{Sherpa} \cite{Sherpa01}, and \textsc{SPEX} \citealt{spex}). However, it is now clear that the physical processes at play produce complex datasets where dimensions are intricately interlinked, necessitating joint modelling of multiple dimensions simultaneously. Notable examples include reverberation mapping and quasi-periodic oscillation (QPO) phase-resolved spectroscopy, which produce both energy and time-dependent signatures (e.g. \citealt{Stevens16}, \citealt{Ingram16}, \citealt{Mastroserio18}, \citealt{Nathan22}); in the near future, polarimetric timing will provide similar challenges, as the data will be a function of both the timescale and modulation angle (\citealt{Ingram15}, \citealt{Ewing25}). Support for these multi-dimensional datasets and models in existing libraries is very limited. At the same time, inferring the model parameters using these multi-dimensional datasets and models is a complex task often not well suited to standard approaches to inference and sampling: parameter correlations and multiple peaks in posterior probability distributions often make the posterior slow to sample, or lead to biased posteriors (e.g. \citealt{Reynolds12}, \citealt{Parker19}, \citealt{Lucchini23}, \citealt{Nathan24} \citealt{Albert25})). To accommodate analyses of X-ray data that involve two or more dimensions, and that robustly produce physically meaningful parameter distributions, the community urgently needs a flexible, modular software framework that enables a wide range of analyses depending on the source under study and the specific scientific question.

The growing popularity of Python over the last decades has had a major impact on scientific software development, availability, and practices. Releasing and contributing to open source packages is increasingly popular among researchers, and the coding and documentation standards in the astronomy community have seen very significant improvement. Owing to these improvements, analysis pipelines can easily be built within the same Python environment by combining features and functionality from different specialised packages. 

Indeed, recent years have seen the release of a wide variety of Python-based packages designed for specific use cases in X-ray astronomy. \textsc{X-PSI} \footnote{\url{https://github.com/xpsi-group/xpsi}} \citep{xpsi} is a library dedicated to pulse profile modelling of neutron star sources. \textsc{Ixpeobssim} \footnote{\url{https://github.com/lucabaldini/ixpeobssim}} \citep{ixpeobssim} is dedicated to spectral-polarimetry analysis of IXPE (Imaging X-ray Polarimetry Explorer) data. \textsc{Sushi} implements a machine-learning-based algorithm in order to separate individual source components in spectral imaging \footnote{\url{https://github.com/JMLascar/SUSHI}} \textsc{Jaxspec} implements the functionality of existing spectral fitting software in a format compatible with machine learning \footnote{\url{https://github.com/renecotyfanboy/jaxspec}} \citep{jaxspec}. \textsc{Stingray} is a generic library for producing and analysing standard spectral-timing data products \footnote{\url{https://github.com/StingraySoftware/stingray}} (\citealt{Stingray1} \citealt{Stingray2}). \textsc{BXA} (Bayesian X-ray Analysis) implements a Bayesian inference framework for standard spectral analysis \footnote{\url{https://github.com/JohannesBuchner/BXA}}  \citep{bxa}. Each of these packages is tailored to specific use cases and/or observables. At the same time, each of these libraries comes with their own limitations: libraries such as X-PSI are specialised for specific scientific use cases and do not easily translate to others. Libraries such as \textsc{JAXSPEC} and \textsc{IXPEOBSSIM} and data analysis environments such as \textsc{XSPEC} are focused on specific data modalities such as spectral or spectro-polarimetric data, which often makes the process of modelling other data modalities (e.g. timing) anywhere from cumbersome to impossible. Sampling libraries such as \textsc{BXA} are often also locked into specific analysis libraries that focus on particular data modalities, and general-purpose inference libraries, in turn, leave the user with the substantial task of assembling a framework for dealing with data, instrument responses, and models themselves. Finally, libraries such as \textsc{stingray} provide generic, flexible analysis pipelines, but to date, no end-to-end solution exists for modelling X-ray data beyond one-dimensional or --in limited cases-- two-dimensional applications. As our data increase in dimensionality and complexity, and as our numerical models become increasingly capable of reproducing multi-dimensional data, so grows our urgent need for a flexible software solution that enables the analysis of these high-dimensional data products and simulations. 

In this paper we detail the first release of \textsc{nDspec}, an open source Python library designed to enable modelling multi-dimensional X-ray data while interfacing with existing Python libraries. The goal of this library is to provide a flexible, modular framework that allows users to easily perform inference on multi-dimensional datasets, and interface with state-of-the-art libraries to enable robust inference of physical quantities. The library's core design philosophy centres on flexibility and extensibility: its functionality and classes are designed with extensibility to higher dimensions in mind. Similarly, the library provides a range of general-purpose functions and classes that enable direct connections to existing optimisation and sampling packages such as \textsc{LMFIT} or \textsc{emcee}. Unlike existing libraries such as \textsc{XSPEC} and \textsc{ISIS}, which provide a data analysis environment, \textsc{nDspec} is designed to provide building blocks that can be flexibly combined into bespoke data analyses tailored towards the specific science case and dimensionality of the data in question. To avoid re-implementation of standard techniques, the library builds upon foundational functionality that exists in well-tested packages such as input and output operations implemented in \textsc{astropy} and Fourier techniques implemented in \textsc{stingray}, and implements interface layers to external models as well as sampling packages.

The initial release detailed here is focused on presenting the overall framework, as well as a use case that is currently challenging to address with existing packages: modelling two-dimensional spectral-timing data produced through Fourier techniques. This is particularly relevant given that our ability to model spectral-timing data with libraries designed for one-dimensional data has not kept pace with the quality of that data, and thus that presents a major bottleneck to our analysis efforts. While spectral polarimetry and polarimetric timing are not yet supported, this functionality is currently in active development and will be the subject of a future release. The paper is structured as follows. In Sec.\ref{sec:core} we describe the core philosophy and functionality of nDspec, and describe how the software handles response matrices, Fourier products, and statistical inference. In Sec.\ref{sec:demo} we demonstrate the functionality of the software by modelling spectral, timing, and spectral-timing data from a typical NICER observation of an accreting black hole X-ray binary. In Sec.\ref{sec:development} we detail our plans for future features in the library, and draw our conclusions. Finally, in Appendix.\ref{sec:appendix} we report the posterior distributions for all the models used in the paper. 

\section{nDspec core functionality}
\label{sec:core}

A key goal of (X-ray) data analysis is to compare data to theoretical (analytical or numerical) models. Current X-ray spectral fitting packages perform model fitting through forward folding: a spectral model in physical units (e.g. flux density as a function of photon energy) is convolved with a response matrix. The response matrix quantifies how the flux is converted into count rates as a function of instrument channel, with each channel nominally corresponding to a photon energy bin (although this mapping is limited by energy resolution as well as complex physical effects within the instrument). Effectively, one takes a physical model and performs a mathematical operation to turn the physical model into units that the data are recorded in, such as photon counts per second. The reason for performing this particular operation is that the response matrix of the instrument is not invertible, and as a result the reverse operation (converting a count rate detected by an instrument into a unique measurement of the flux as a function of energy) is not possible \citep{Buchner23}. The data and the model transformed into detector space are then compared statistically using a likelihood function, and estimates for parameters are usually found via optimisation (i.e.~Maximum Likelihood Estimation) or sampling (Bayesian inference). Most existing packages largely abstract both the technical forward-folding operations and the statistical methods away from the user. This yields a relatively simple interface for the user, at the cost of flexibility: because these software packages prioritise simplicity, they internally choose either optimisation or sampling algorithms as well as standard practices or settings for these algorithms.

\begin{figure}
\centering
\includegraphics[width=\columnwidth]{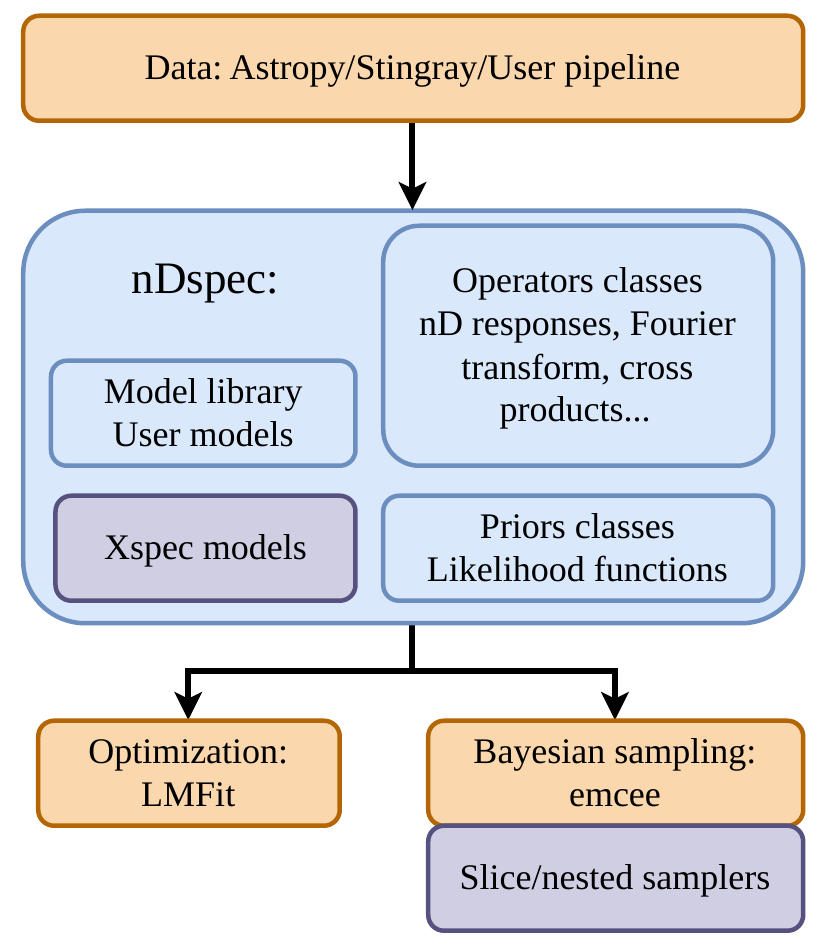}
\caption{Schematic of the workflow enabled by nDspec. Orange boxes indicate existing libraries and functionality, and blue boxes indicate functionality provided by nDspec to tie these libraries together. Purple boxes denote straightforward additions to the package that will be included in the short term.}
\label{fig:schematic}
\end{figure}

\textsc{nDspec} extends this capability to multiple dimensions, by defining multiple `operator' classes. Operators define all operations to be done to the model in order to compare it to the data. For example, one might define a model as flux as a function of photon energy and time, and use one \textsc{nDspec} operator to fold the instrument response, another to compute the Fourier transform, and a final operator to derive model time lags in as a function of energy channel in a given Fourier frequency range \citep{Uttley14}. This formalism can also be extended to X-ray polarimetry - for example, when converting from a polarisation model to its corresponding modulation curve \citep{Ingram22Review}. The fundamental philosophy of \textsc{nDspec}, and a key difference from existing, one-size-fits-all analysis packages, is to provide both stand-alone classes for each operator that users can implement in their own workflows, and a set of classes capable of optimisation and inference that handle the appropriate operators internally for the most common use cases (such as modelling lags). We envision users being able to pick and choose individual features of \textsc{nDspec} as appropriate for their use case, and incorporating these specific features in their analysis.

Similarly, at the other end, \textsc{nDspec} provides flexible interfaces to existing sampling and optimisation algorithms, with straightforward pathways for extending these interfaces to new algorithms and libraries as they are developed, or to specialised techniques appropriate for a particular scientific use case. In this way, \textsc{nDspec} enables bespoke workflows tailored to specific scientific datasets, which enable more robust and precise scientific results compared to one-size-fits-all solutions. A schematic of how nDspec ties together existing libraries with its own functionality is shown in Fig.\ref{fig:schematic}.

\begin{figure*}
\centering
\includegraphics[width=0.5\columnwidth]{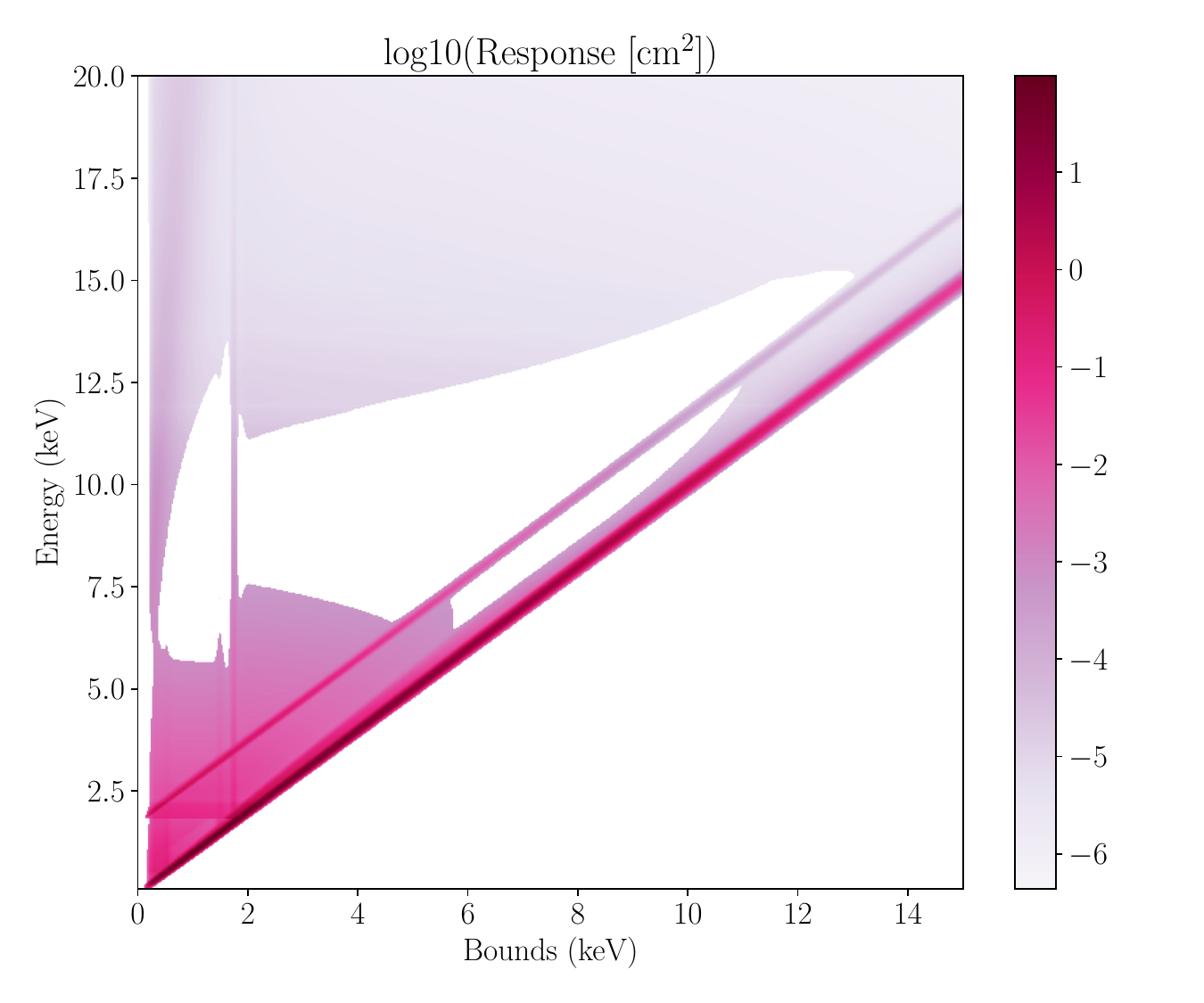}
\includegraphics[width=0.5\columnwidth]{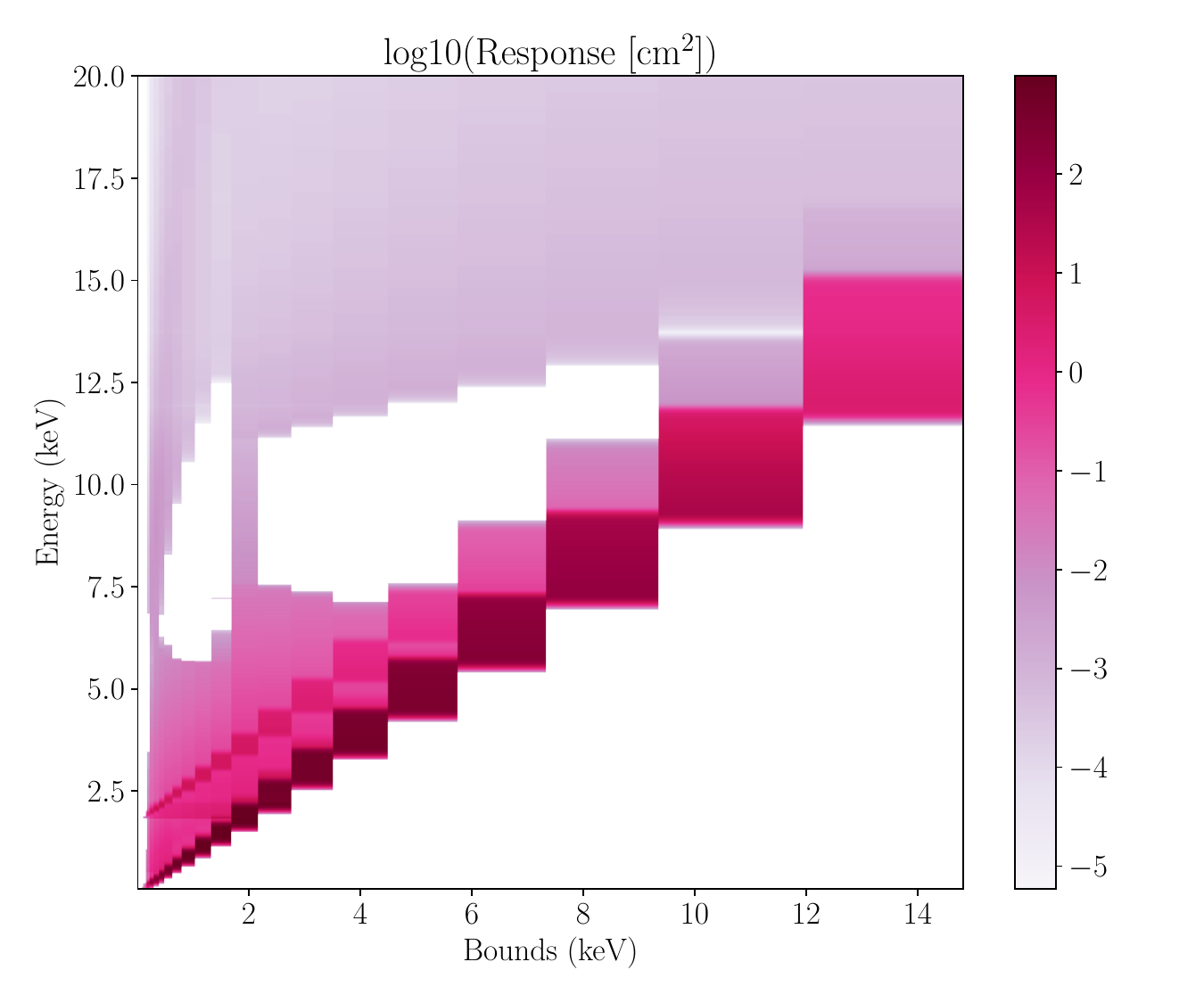}
\includegraphics[width=0.5\columnwidth]{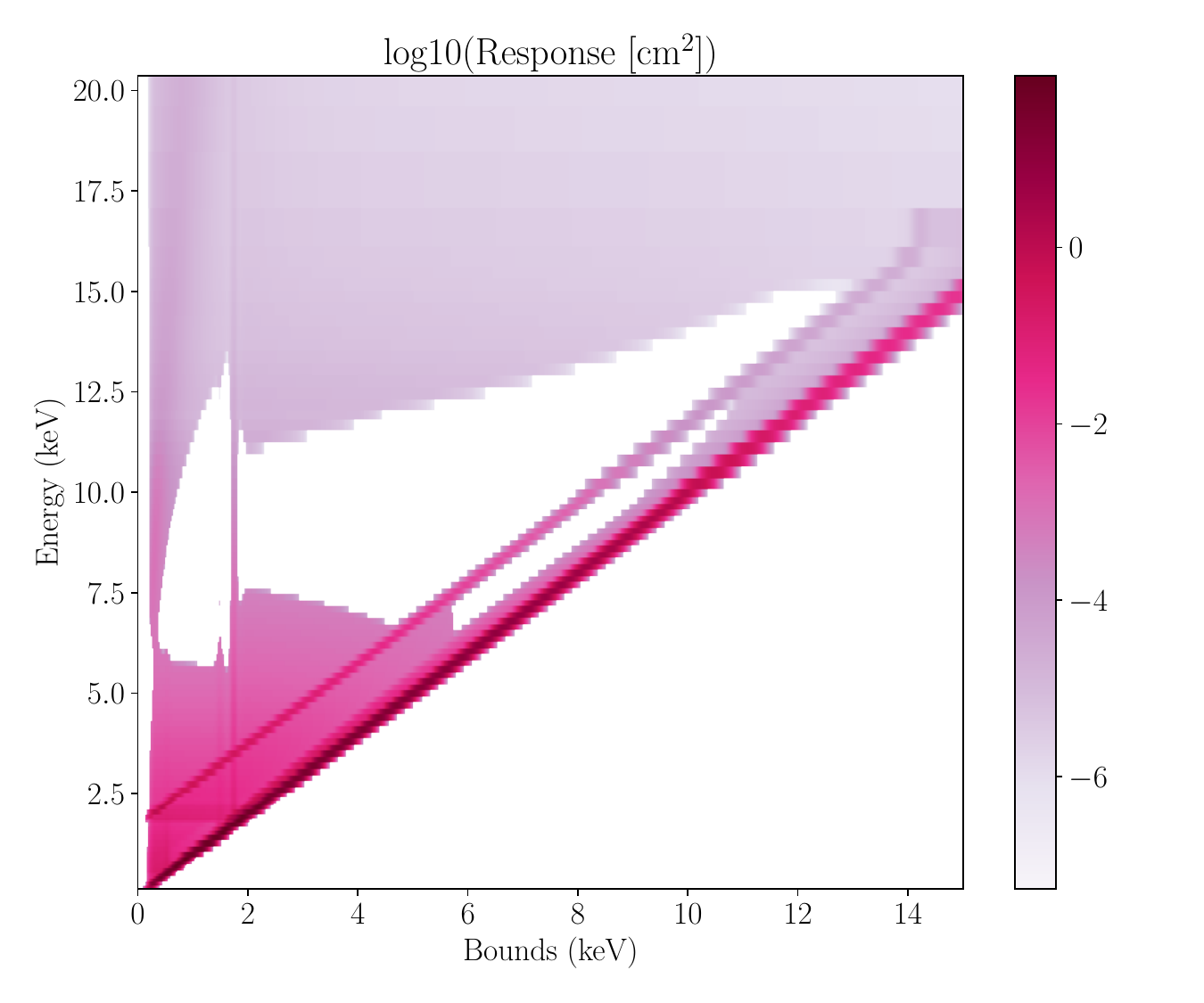}
\includegraphics[width=0.5\columnwidth]{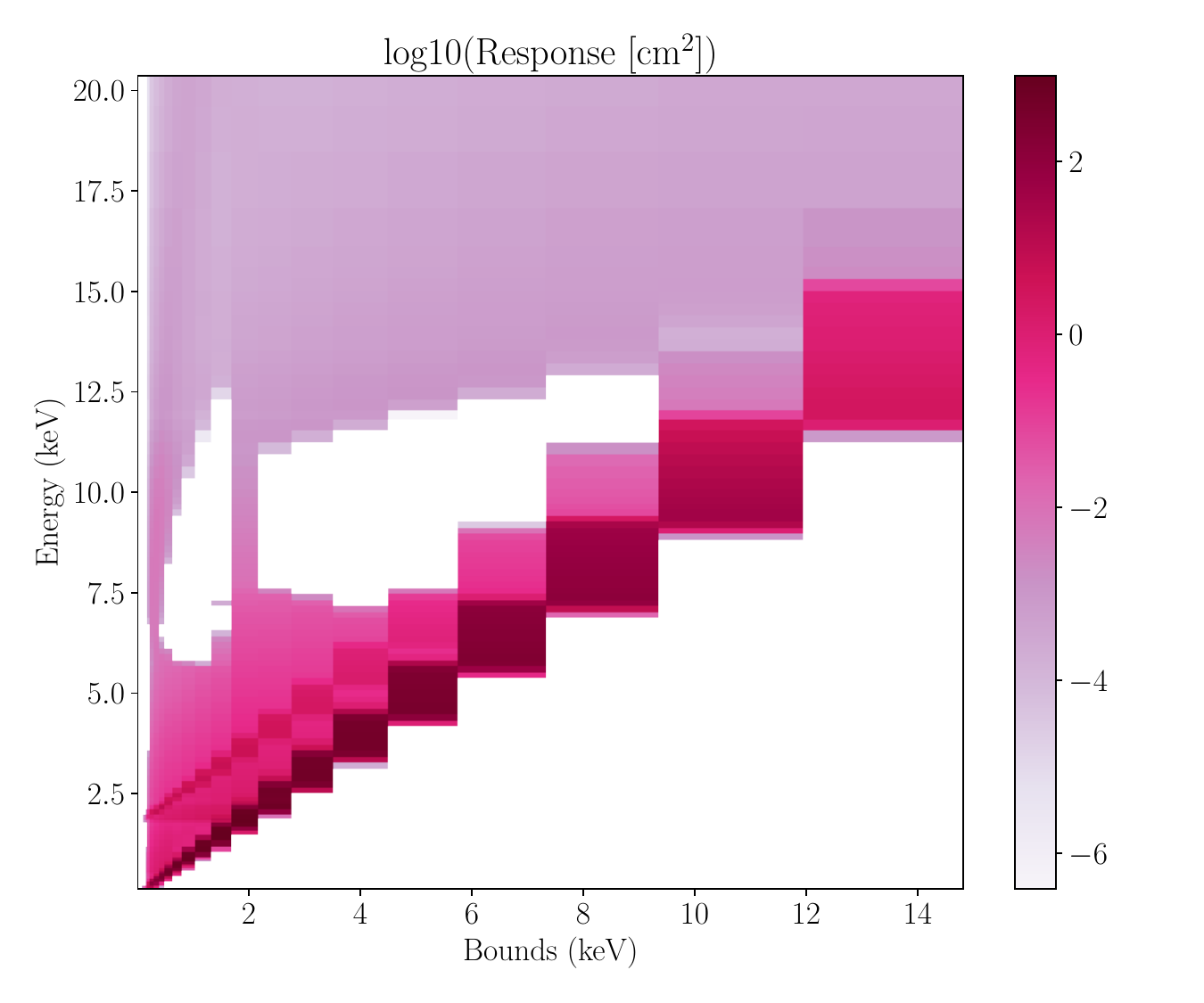}
\caption{NICER instrument response with different binning. Top left: the response in its native resolution, 3451 physical energy bins$\times$1501 channels (roughly containing 100 channels per keV). Top right: the response re-binned over channels down to 30 geometrically spaced energy channels. Bottom left: the response re-binned over energies by a factor 30. Bottom right: the response re-binned over energies by a factor of 30, and over channels to 30 geometrically spaced channels. }
\label{fig:response_bin}
\end{figure*}

\begin{figure}
\centering
\includegraphics[width=\columnwidth]{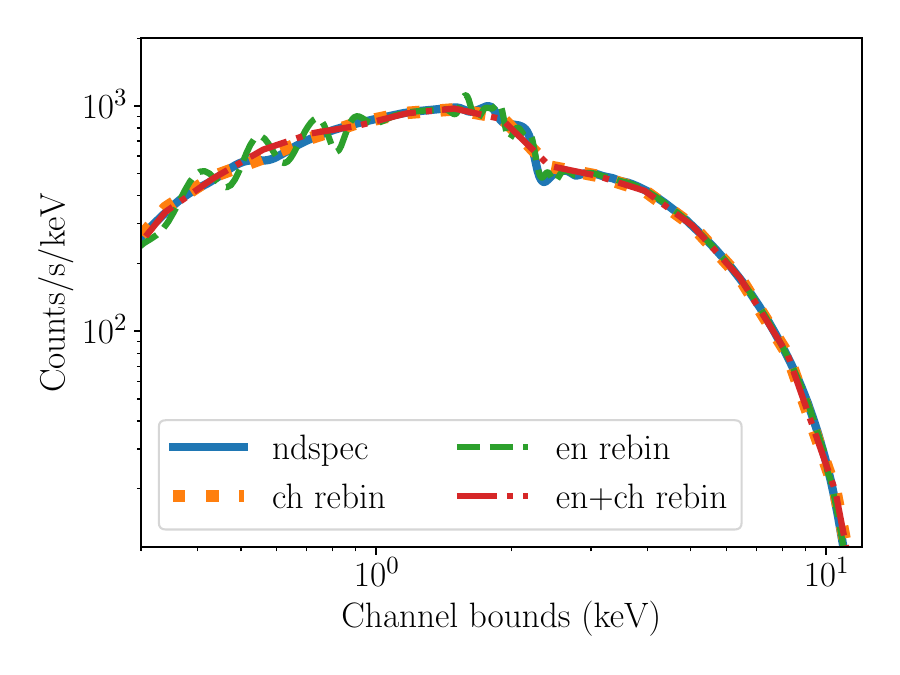}
\caption{Comparison of nDspec folding a constant model. The blue, orange, green and red lines correspond to the top left, top right, bottom left and bottom right panels, respectively. As shown by the green line, re-binning over energies introduces distortions in the spectrum if the number of energy bins is not larger than the number of channels. On the other hand, re-binning over channels (orange line) or over both quantities (red line) does not significantly alter the output spectrum.}
\label{fig:folding_schemes}
\end{figure}

\begin{figure}
\centering
\includegraphics[width=\columnwidth]{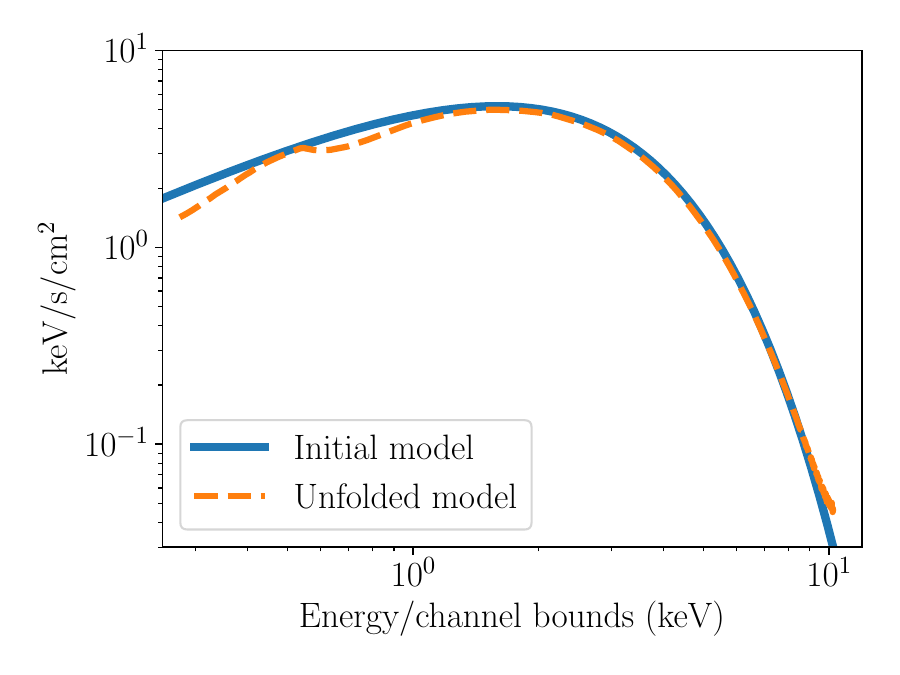}
\caption{Comparison of a physical model before folding with the Swift?XRT response matrix (blue line), and of the same model folded and then un-folded in \textsc{nDspec} (orange line). The two agree over most of the energy range, but there is a noticeable discrepancy at energies below $\approx 1$ keV.}
\label{fig:nDspec_unfold}
\end{figure}

\subsection{The \texttt{ResponseMatrix} class}
\label{sec:response}

Instrument response matrices in nDspec are read through the \textsc{Astropy} \textsc{io} package (\citealt{astropy13}, \citealt{astropy18}, \citealt{astropy22}) and stored in two-dimensional \textsc{NumPy} \citep{numpy} arrays; in this format, the process of folding a (one- or two-) dimensional models with \textsc{NumPy} can be carried out as a simple matrix multiplication. With this definition, the results of folding a model through the instrument response is identical to \textsc{Xspec}. The unit tests included in \textsc{nDspec} ensure that when folding instrument responses, the two packages agree to at least a numerical accuracy of one in a million. The class additionally contains methods to plot the response (and, if it is defined separately, the effective area), to optimise the folding operation by re-binning the response over both channels and energies, and a method to produce `unfolded' data.

Response matrices generally take a spectrum in physical units as a function of energy, and redistribute the photons that arrived at the detector into the relevant detection channel. Thus, re-binning is possible in two dimensions: firstly, more commonly one may re-bin the output channels (i.e.~columns of the response matrix) to facilitate further analysis with the spectrum. Secondly, one may also wish to re-bin the response matrix in the input dimension (energies; rows in the response matrix), particularly when calculations with the full response would demand a high computational load. 

Re-binning data (and responses) in channels is a typical operation in X-ray spectral fitting. In this section, we discuss its applications to spectral-timing analysis, which is typically conducted over much coarser energy resolutions than standard spectral analysis. For example, even high throughput instruments like NICER allow the use of $\approx$tens of channels at most with sufficient signal to noise ratios (e.g. \citealt{Stevens18}, \citealt{Wang21}, \citealt{Bellavita22}). The \texttt{ResponseMatrix} class allows users to re-bin a response matrix over an (almost) arbitrary user-defined grid of channels: in order to avoid potential issues with instrument gain, the user-defined grid is shifted automatically to match the existing bin edges, but otherwise any user-specified binning is valid. This means that when re-binning channels are always combined, and never split. This implementation is essentially identical to the \textsc{HEASOFT} \texttt{rbnrmf} tool, and is included in nDspec to allow users to explore the effect of the channel resolution in their analysis, without the need for external tools. 

Re-binning a given response over (input) energies is generally not recommended as it may introduce artefacts into the response that may propagate biases into the scientific results, as we will show below. It is, however formally supported by the \textsc{HEASOFT} \textsc{ftools}, and may be desirable in some cases in order to optimise model runtime. Two common examples are if the model computations are extremely fast, as can be the case for models computed from trained neural networks, (e.g. \citealt{Ricketts25}), or if the model runtime is long despite the data energy resolution being low, as is the case for spectral timing models that are also a function of time and/or Fourier frequency. The \texttt{ResponseMatrix} class supports this operation, but we urge the user to proceed with caution. Unlike re-binning over channels, users can only re-bin in energy by integer grouping: for instance, re-binning by a factor of ten simply means grouping ten successive energy bins into a single new one. Again, this implementation is essentially identical to that in \textsc{HEASOFT}.

To illustrate the challenges with re-binning over input energies, we show a comparison of \textsc{nDspec} \texttt{ResponseMatrix} instances which have different binning schemes in Fig.\ref{fig:response_bin}. The top left panel shows the base NICER response, consisting of 3451 input photon energy bins and 1501 output energy channels, with channels of width $\approx$0.01 keV. The top right panel shows the same instrument response, re-binned to a much coarser channel grid of 30 geometrically spaced channels. This kind of channel binning is more typical of spectral-timing analysis. The bottom left panel shows the response binned over energies by a factor of 30. Finally, the bottom right shows the response binned over both energy and channels.

The impact of the different binning schemes is shown in Fig.\ref{fig:folding_schemes}, using a constant model for simplicity. The most accurate result is that computed from the un-binned response, shown in the blue line. Re-binning over channels exclusively (orange dotted line) does not introduce any distortions in the output, and simply results in the instrumental features being smeared out. On the other hand, re-binning in energy alone (green dashed line) shows very sharp artificial features around and below $\approx$2 keV; these features are a textbook example of why re-binning in energy can be dangerous and is (rightfully so) very rarely done. On the other hand, re-binning in both energy and channels (red dash double dotted line) actually causes these features to average out over the coarse channel grid and disappear. 

To conclude, for most scientific applications, re-binning over input energies may be acceptable if re-binning is performed over both input energies and output channels in order to average out artefacts, as is usually the case for typical spectral-timing applications, where coarse channel grids are common. This re-binning (in both energy and channels) reduces the computational cost of folding the response by a factor $\approx\sqrt{N}$, where $N$ is the re-binning factor used to re-bin the response in either dimension. Depending on the model computation and the required model energy resolution, the trade-off between precision and computational speed may be acceptable. 

The final functionality provided by the \texttt{ResponseMatrix} class is to `unfold' data, which is the operation of converting the observed count rate in each channel to a physical flux in units of energy per unit time and area. Because, as mentioned above, the response matrix is not strictly invertible, the deconvolution of the intrinsic spectrum with the instrument response necessarily requires approximations that make the resulting unfolded spectrum not reliable for drawing scientific conclusions, since any approach to unfolding introduces deviations from the true flux emitted by the source (e.g. \citealt{Buchner23}). In \textsc{nDspec} we utilise the same implementation as \cite{isis}, and define an unfolded spectrum $F(h)$ as
\begin{equation}
    F(h) = \frac{C(h)}{t\int_{\Delta E(h)}\mathcal{R}(h,E)dE},
    \label{eq:unfold}
\end{equation}
where $C(h)$ is the background-subtracted observed spectrum as a function of energy channel $h$, $t$ is the exposure time, $\mathcal{R}(h,E)$ is the instrument response matrix, and $\Delta E(h)$ indicates that the integration is carried out over energies $E$ that are read in channel $h$. When unfolding, eq.\ref{eq:unfold} is applied to both model and data - effectively, it works as a phenomenological re-normalisation designed primarily to aid users in visualising their data. Compared to the traditional \textsc{Xspec} implementation, the advantage of this definition of an unfolded spectrum is that it is entirely model independent. The downside however is that the true flux is not reconstructed perfectly, even in ideal cases when the model exactly reproduces the data. We show this in Fig.\ref{fig:nDspec_unfold}, where we compare a $1$ keV black body, with the same model folded and then un-folded through the Swift/XRT response. While at energies above $\approx1$ keV the two are essentially identical, the unfolded flux deviates from the original spectrum at low energies. This deviation is inherent to the process of unfolding, regardless of instrument, and is a fundamental reason for why one should forward-model the observed data, rather than attempting to convert the data into the units of the model. We strongly encourage users to read \cite{unfold} for more details; the authors discuss in-depth the caveats and potential issues of unfolding X-ray data, and demonstrate that Eq.\ref{eq:unfold} is robust enough for preliminary analyses and data visualisation. Finally, we note that $F(h)$ need not be just a time-averaged spectrum; one can apply the same operation to, for example, an rms or phase-resolved spectrum as well, provided that the data is proportional to a count rate per channel.
 
\subsection{The \texttt{PowerSpectrum} and \texttt{CrossSpectrum} classes}
\label{sec:timing}

\textsc{nDspec} handles spectral-timing model calculations through two classes, which handle power spectra and cross spectra respectively. In addition to class-specific features, both classes allow the conversion of time-dependent model in the Fourier domain using two different implementations of the Fourier transform. The first option is to use the common \textsc{FFTW3} library, through its Python wrapper \textsc{pyFFTW}. The main downside of this option is that users are limited to linearly spaced time (and therefore Fourier frequency) grids. Alternatively, the library includes an implementation of the sinc function decomposition described in \cite{Uttley25}, which allows users to define arbitrarily spaced (e.g., logarithmically) time and Fourier grids, potentially gaining computational speed, at the cost of some computational accuracy. 

The power spectrum class \texttt{PowerSpectrum} can be defined either starting from an array of times or frequencies, and models can be defined in the time domain and Fourier transformed or directly in the Fourier domain. It is a fairly simple class, mostly intended for use with models of the cross spectrum that require some assumption about the power spectrum as well.

The cross spectrum class \texttt{CrossSpectrum} is more complicated, as the latter is a complex (rather than real) quantity that is function of both energy (through the choice of subject band(s) to compare to a reference band) and Fourier frequency; additionally, users may be interested in modelling exclusively the time lags encoded in the phase of the cross spectrum. The main function of \texttt{CrossSpectrum} is to allow users to define models both in the time and Fourier domains, convert from the former to the latter, and to output standard spectral-timing products of their choice (such as a lag vs frequency, or modulus vs energy). These functions in turn enable users to greatly simplify defining and combining models compared to existing software, as unlike existing software a) a single model or parameter instance is sufficient to cover every bin in both energy and frequency, and b) combining model components is taken care of by the software no matter the coordinates in which the data and model are defined.

In the case of time domain models, \textsc{nDspec} makes use of the linear impulse response function (IRF) formalism. This formalism has been used commonly throughout the literature, and is particularly useful for reverberation models (\citealt{Campana95}, \citealt{Wilkins16}, \citealt{Chainakun17}, \citealt{Mastroserio18}, \citealt{Lucchini23}, \citealt{Uttley25}). In this formalism, the time-dependent flux in a given energy band $f(E,t)$ can be written as
\begin{equation}
    f(E,t) = s(t) \circledast g(E,t),
    \label{eq:linear_irf}
\end{equation}
where the impulse response function $g(E,t)$ quantifies the energy and time-dependent response of the system to a $\delta$ perturbation (like a flash of mono-energetic photons). $s(t)$ represents some mechanism responsible for driving the variability - for instance, this might encode fluctuations in the accretion rate that propagate through an accretion flow (\citealt{Kotov01}, \citealt{Arevalo06}, \citealt{Rapisarda14}). The assumption of linearity means that Eq.\ref{eq:linear_irf} is either explicitly linear in time, or can be linearised through a Taylor expansion in which the second and higher orders are negligibly small. 

In practice, Eq.\ref{eq:linear_irf} is rarely computed explicitly due to the computational cost involved in performing a convolution, and the need to average over the noise process driving the variability. Instead, it is convenient to invoke the convolution theorem, which states that a convolution in the time domain corresponds to a (computationally trivial) product in the Fourier domain. The Fourier transform of Eq.\ref{eq:linear_irf} then is
\begin{equation}
    F(E,\nu) = S(\nu)G(E,\nu),\end{equation}
where $G(E,\nu)$ is the Fourier transform of $g(E,t)$, and is referred to as the transfer function; $S(\nu)$ is the Fourier transform of $s(t)$. From $F(E,\nu)$, one can calculate the cross spectrum $C(E,\nu)$ (averaged over many realisations of the noise process, \citealt{Uttley14}) between any given energy band $E$ and a reference band $E_{ref}$,\begin{align}
    C(E,\nu)  & = F(E,\nu)F^{*}(E_{ref},\nu) = S(\nu)G(E,\nu)S^{*}(\nu)G^{*}(E_{ref},\nu)  \nonumber \\
   & =  S^{2}(\nu)G(E,\nu)G^{*}(E_{ref},\nu),
    \label{eq:cross_spec}
\end{align}
where $S^{2}(\nu)$ is, by definition, the power spectrum of whatever mechanism is responsible for driving the variability $P(\nu)$. Note that the only energy dependence of the cross spectrum (and therefore its derived products, such as lag spectra) is encoded in the transfer function $G(E,\nu)$; $P(\nu)$ only acts as a frequency-dependent weighting term. \textsc{nDspec} supports two-dimensional models for the impulse response $g(E,t)$, the transfer function $S(E,\nu)$, or the cross spectrum $C(E,\nu)$; the latter case allows users to use models that do not make use of the impulse response formalism.

Regardless of the model type, the \texttt{CrossSpectrum} class includes methods to compute both $G(E,\nu)$ and $C(E,\nu)$, and to convert the latter into lags alone, or a full cross spectrum in either Cartesian or polar coordinates, expressed as a function of either frequency or energy. In the case of energy-dependent products, the code averages over Fourier frequency identically to \cite{Mastroserio20}. We note that the assumption of linearity implies that the intrinsic coherence is unity at every Fourier frequency and energy, and higher order Fourier products (such as the bi-spectrum or cross-bi-spectrum) are zero. This is not the case when higher order terms in the Taylor expansion are not negligible \citep{Hardin86,bendat2010}. A full treatment of non-linear impulse response functions is beyond the scope of this paper. 

\subsection{Scientific inference with nDspec}

\textsc{nDspec} implements a range of classes to facilitate statistical inference with (multi-dimensional) X-ray data. At a high level, this interface is designed to help users define and combine models for various spectral-timing products, compare these models to data, and derive meaningful parameter distributions for the underlying physical process. Two routes for inference currently exist, though \textsc{nDspec}'s modular framework enables straightforward integration of additional algorithms and libraries as they are developed and released. Current algorithms include optimisation to obtain best-fit parameters, implemented in the optimisation classes, and Bayesian inference and Markov Chain Monte Carlo (MCMC) posterior estimation. We recommend using the optimisation interface primarily for exploratory purposes and to estimate a starting position for MCMC posterior estimation, and urge the user to make use of the MCMC interface to obtain robust posterior distributions for parameters relevant to the physical conclusions of any data analysis project.

\subsubsection{The optimisation classes}
\label{sec:optimization}

\textsc{nDspec} handles the process of optimisation through a set of specialised classes, which allow users to fit specific types of data (such as time-averaged spectra, or lags) by combining the \textsc{LMFIT} optimisation library \citep{lmfit} with the appropriate operators. For example, spectral models can be defined in traditional units of flux per unit energy, after which the optimisation class automatically performs the convolution with the instrument response through an instance of the \texttt{ResponseMatrix} class. In similar fashion, when fitting lag spectra using an impulse response function, the code performs the operations described in Sec.\ref{sec:timing} as well as the instrument response convolution. Regardless of the operator(s) used, the outcome is an array with the same dimension and units as the data, allowing the immediate computation of a log-likelihood used for optimisation.

The alpha release includes a small library of phenomenological models (described in Appendix \ref{sec:models}) in the form of Python functions. In the short term, we assume that users will be able to either write their own models in Python, or wrap their code in other languages using a Python interface like \textsc{pybind} or \textsc{f2py}. We plan to provide wrappers for the Xspec library of models while avoiding use of any existing software (such as \textsc{PyXspec}) in the near future. Model functions are then turned into \textsc{LMFIT} \texttt{Model} objects, which allows users to handle parameters and their values, combine model components, and so on, identically to existing spectral analysis software regardless of the type of data they are interested in. 

The functions and methods for loading data are also designed to give users flexibility. \textsc{NumPy} arrays can be used to pass data for all types of datasets currently supported. Additionally, each specific optimisation class has a dedicated way of interfacing with existing infrastructure, from \textsc{Stingray} objects to OGIP-compatible files depending on the fitter type.

The \texttt{FitCrossSpectrum} class includes additional functionality to be included in model evaluations. First, as discussed in Appendix E of \cite{Mastroserio18}, calibration uncertainties in the instrument response can introduce a small phase in the energy-dependent cross spectrum, which may additionally be frequency-dependent. This occurs because if the energy of the photons is not reconstructed perfectly (which is never the case) then some photons that were assigned to channels in the reference band will have an energy outside of those boundaries, and vice versa. \textsc{nDspec} allows users to optionally correct for this effect by including a small phase in each Fourier frequency bin as a free parameter, regardless of the model used. This is identical to the $\Phi_A(\nu)$ parameter included in \textsl{reltrans}. Additionally, models for the energy-dependent cross spectrum derived from impulse or transfer functions require frequency-averaging over an assumed power spectrum ($P(\nu)$ in eq.\ref{eq:cross_spec}), which is not always known. An incorrect assumption on the form of $P(\nu)$ will introduce (small) frequency-dependent deviations of the modulus from its correct form; this correction can also be optionally accounted for in \textsc{nDspec} by including a multiplicative factor on the modulus of the model cross spectrum.

The current version only supports data with Gaussian uncertainties through the $\chi^2$ statistic; additional cases including a Poisson likelihood (Cash statistic; \citealt{Cash79}) and a $\chi_2$ likelihood (Whittle statistics; \citealt{Vaughan10}) will be included in a future release. By default, the code uses the standard Levenberg-Marquardt minimisation, however \textsc{LMFIT} includes implementations of roughly a dozen additional optimisation methods. A performance comparison of these methods is beyond the scope of this paper, but users are encouraged to test whether their particular data is better suited to a different algorithm. In the current version, estimating uncertainties on the best-fit parameters via Fisher information is not yet implemented, and we encourage users to use Bayesian sampling instead.

\subsubsection{The MCMC interface}
\label{sec:emcee}

The classes described perform optimisation: they find the combination of model parameters which (ideally globally, but more likely locally) result in the lowest fit statistic, defined by some convergence criterion. This result will only coincide with the maximum likelihood in the case of a global maximum, and makes no robust statements about the uncertainties in the parameter, which are crucial for realistic modelling applications. For this use case, \textsc{nDspec} enables Bayesian inference to compute posterior probability densities of the parameters via various sampling algorithms. In the alpha release, we implemented a set of functions and classes to facilitate interfacing \textsc{nDspec} objects with the common \textsc{emcee} Python library \citep{emcee} for Markov Chain Monte Carlo (MCMC) sampling.

Given an MCMC sample, we provide a function for easily visualising the results. In particular, it allows users to compute the auto-correlation lengths for each parameter and plots trace plots for all walkers, walker acceptance ratios, and corner plots for all parameters; by default, the latter plots are only displayed if the chain has been run long enough (defined as at least 50 times the integrated auto-correlation length of each parameter), although users can bypass this setting (for example, to monitor a chain while it is still running). By tying ease of visualisation of a chain with its convergence, we hope to both allow users to confirm the robustness of their findings, and encourage good statistical practice within the community.

\section{A demonstration: Characterisation of a NICER observation}
\label{sec:demo}

In this section, we analyse a typical NICER observation of a black hole X-ray binary, in order to showcase a typical workflow in \textsc{nDspec}. The observation we chose is OBSID 1200120106, a canonical bright hard state of the transient MAXI J1820+070. We note that while this analysis is technically possible with existing packages, it is typically cumbersome and slow to do. We also emphasise that reproducing functionality existing in other packages is only a starting point, and \textsc{nDspec}'s true power lies in its modular approach, which enables rapid integration of additional dimensions and sampling algorithms. We begin by modelling the power spectrum and time-averaged spectrum, and then use information from these fits to model the energy and frequency dependent cross spectrum. 

\subsection{Data reduction}

We reduce the data with \textsc{Chromie}, our own set of Python scripts which combine the standard NICER tools in \textsc{HEASOFT} (at the time of reducing the data, we used version 6.32.1) with \textsc{Stingray} (version 2.1 at the time of writing) to produce timing and spectral-timing products. The scripts are publicly available at \url{https://github.com/matteolucchini1/Chromie}. 

We produce level 2 event files using \texttt{nicerl2}, using the default screening criteria. From the level 2 cleaned event file, we extract instrument responses and a time-averaged spectrum using \texttt{nicerl3-sp}, ignoring detectors 14 and 34 due to their noise and otherwise using the default settings (including applying a $\approx1.5\%$ systematic error to the data). We use the default SCORPEON model for the instrument background; however, due to the extremely high count rate ($\geq 10^4$ counts/s) of the observation and the exploratory nature of the inference discussed here, we do not include an additional background component in our model for simplicity, as it is effectively negligible. Before modelling the data, we re-bin the spectrum identically to \cite{Wang21}.

We use \textsc{Stingray} to calculate the averaged power spectrum from the event file, using segments of $256$s with a time resolution of $1$ms and including events between channels 50 and 1000 (nominally corresponding to 0.5-10 keV). We normalise the power spectrum to fractional rms units. After calculating this averaged power spectrum, we use \textsc{Stingray} to subtract the Poisson noise and re-bin the data geometrically by a factor $f=1.05$. We ignore data above $20$Hz, as the signal variability power in BHXRBs drops off at frequencies above $\approx$tens of Hz.

Finally, we extract energy-dependent lightcurves with \texttt{nicerl3-lc} in 41 geometrically spaced bins between channels 50 to 1000 (corresponding roughly to energies between 0.5 and 10 keV), re-aligning the edges of the channel grid to the NICER response were necessary. We also extract a lightcurve to use as a broad reference band using the same 50-1000 channel range. We use a time resolution of $0.03$s for all the lightcurves. We then define six geometrically spaced Fourier frequency bins between 0.2 and 16 Hz, and once again use \textsc{Stingray} to calculate the energy dependent cross spectrum (modulus, phase, and time lags) in polar coordinates following \cite{Uttley14}, averaging over $5$s segments and utilising absolute rms units. When modelling the cross spectra, we re-bin the response matrix in energy by a factor of $10$ (thus obtaining a matrix with dimensions $346\times40$) in order to speed up model evaluations. Before proceeding we verified that this choice does not introduce any distortions in the model computations as discussed in Sec.\ref{sec:response}. 

\begin{figure}
\centering
\includegraphics[width=\columnwidth]{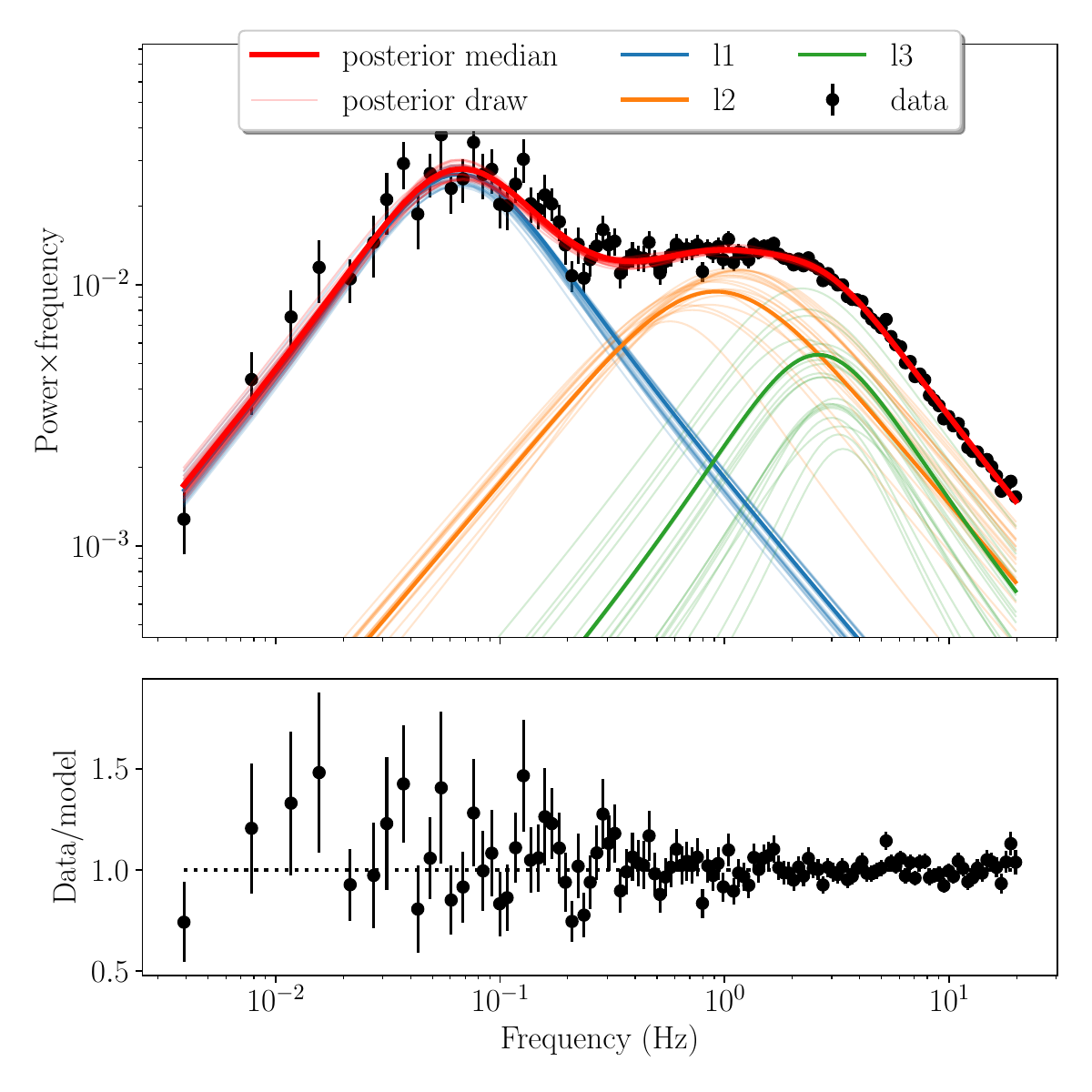}
\caption{Data (black) together with the median of the posterior distribution (solid red line) and random draws from the posterior (pink lines). In addition to the data and total model, we also show the individual Lorentzian components for the posterior median only. While the contribution of the first Lorentzian is easily separated, the second and third overlap. The residuals are calculated using the posterior median.}
\label{fig:psd_fit}
\end{figure}

\begin{table}[]
    \centering
    \caption{Inference results for the power spectrum.}
    \small
    \begin{tabular}{c|c|c|c}
        Parameter & least-$\chi^{2}$ & Posterior & Prior\\ \hline 
        Lorentzian 1 $\nu_{peak}$ (Hz) & $6.4\cdot10^{-2}$ & $6.5^{+0.4}_{-0.3}\cdot10^{-2}$ & [0,0.3] \\
        Lorentzian 1 $Q$ & $0.34$ & $0.31^{+0.07}_{-0.06}$ & [10$^{-5}$,2] \\
        Lorentzian 1 $rms$ & $0.250$ & $0.251^{+0.006}_{-0.007}$ & [10$^{-3}$,1] \\
        Lorentzian 2 $\nu_{peak}$ (Hz) & $0.8$ & $0.8^{+0.2}_{-0.2}$ & [0.1,1.25] \\
        Lorentzian 2 $Q$ & $0.08$ & $0.09^{+0.12}_{-0.06}$ & [10$^{-3}$,1] \\
        Lorentzian 2 $rms$ & $0.160$ & $0.160^{+0.017}_{-0.025}$ & [10$^{-3}$,1] \\
        Lorentzian 3 $\nu_{peak}$ (Hz) & $2.4$ & $2.5^{+0.4}_{-0.3}$ & [1.25,10] \\ 
        Lorentzian 3 $Q$ & $0.33$ & $0.34^{+0.16}_{-0.07}$  & [10$^{-5}$,2] \\
        Lorentzian 3 $rms$ & $0.12$ & $0.12^{+0.03}_{-0.03}$ & [10$^{-3}$,1] \\ 
    \end{tabular}
    \label{tab:psd_pars}
\end{table}

\subsection{The power spectrum}

We model the power spectrum with the canonical prescription of a sum of Lorentzian functions. We find that the power spectrum is well described by three broad Lorentzians, without the need of additional narrow features for QPOs. We first perform a likelihood optimisation using the $\chi^2$ statistic to find a set of best-fit parameters. The fit statistic for this model is $\chi^{2}/\rm{d.o.f.}=110.28/105=1.05$. After finding a plausible minimum, we sample the posterior distribution using \textsc{emcee}, using the optimisation results as a starting position. We use uniform priors for all the parameters, limiting the range of the peak frequencies of each Lorentzian to $[0,0.3]$, $[0.3,1.25]$ and $[1.25,10]$Hz respectively to retain the same order for the individual Lorentzian components across the run. We use 32 walkers and run the sampler for 100000 steps, discarding the first $3000$ steps as burn-in. The resulting walkers are well-mixed and show no correlations. We find that the autocorrelation length of the remaining walkers is $\approx$300 steps for the parameters of the first Lorentzian, and $\approx$1300 for those of the second and third. Based on this autocorrelation length, and thin the chain by a factor of $700$ to produce the posterior distributions. We show the data along with the posterior median and random draws from the posterior in Fig. \ref{fig:psd_fit}. A summary of the best-fitting parameters from the initial optimisation as well as the 68th percentile credible intervals derived from the \textsc{emcee} run is shown in Tab.\ref{tab:psd_pars}. The one- and two-dimensional posterior distributions for all model parameters are shown in Fig.\ref{fig:psd_posterior} of the Appendix.

\begin{figure}
\centering
\includegraphics[width=\columnwidth]{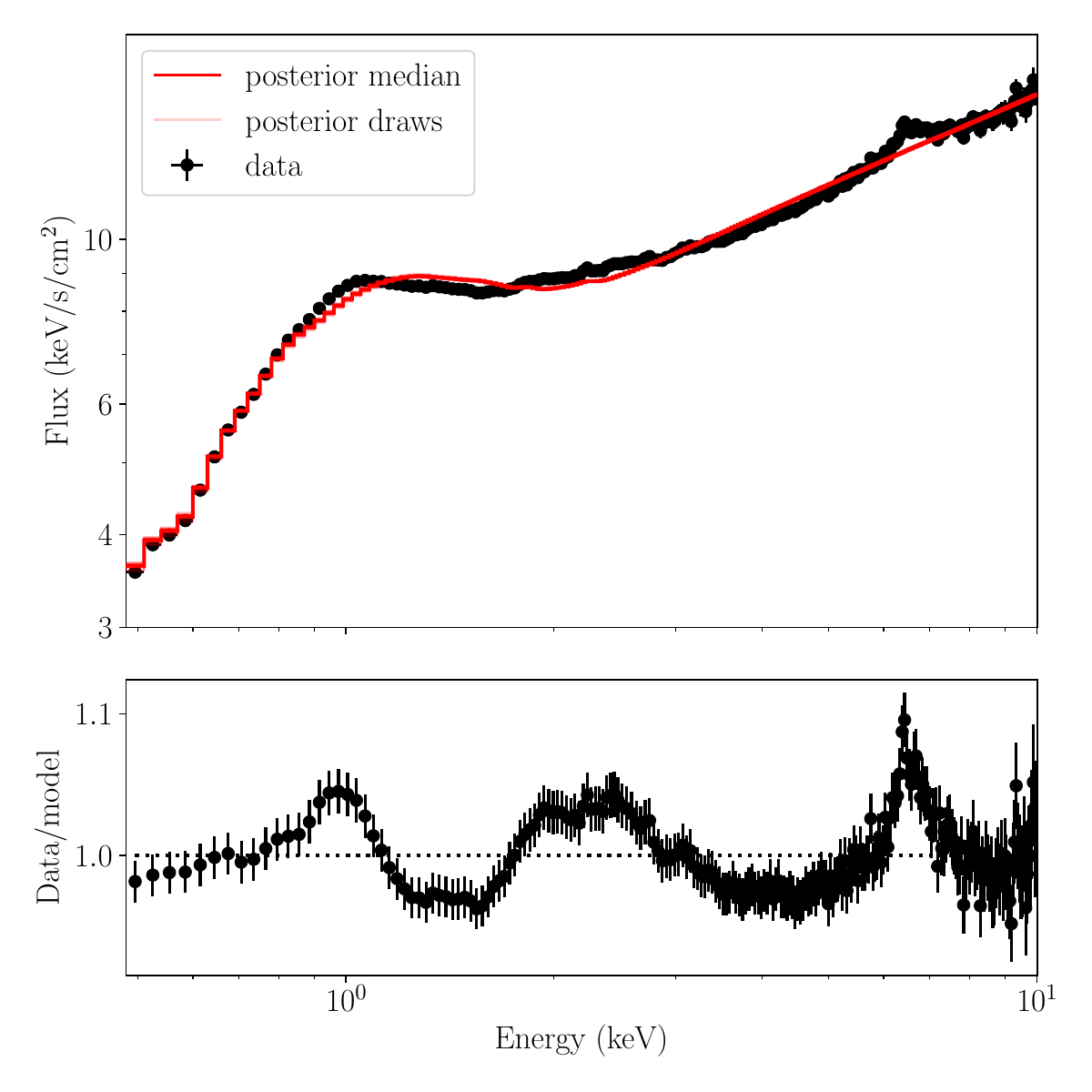}
\caption{Time-averaged spectrum (black) with the median of the posterior distribution (red) and random draws from the posterior (pink). The posteriors are very narrow, such that the posterior draws significantly overlap with each other and the posterior median as to be nearly invisible. The residuals are calculated using the posterior median only. Reflection features clearly visible in the residuals; modelling these is beyond the scope of this work.}
\label{fig:spectrum_fit}
\end{figure}

\begin{table}[]
    \centering
    \caption{Parameter inference for the time-averaged spectrum; for the posteriors, we report the median of the posterior distribution along with the $68$th percentile interval for each one-dimensional posterior.}
    \small 
    \begin{tabular}{c|c|c|c}
        Parameter & least-$\chi^{2}$ & Posterior & Prior \\ \hline 
        $nH$ (cm$^{2}$)& $0.097$ & $0.097^{+0.003}_{-0.003}$ & [0.02,2] \\
        $\Gamma$ & $-1.601$ & $-1.601^{+0.005}_{-0.005}$  & [-2.2,-1.2] \\
        Power-law $norm$ & $6.24$ & $6.24^{+0.05}_{-0.05}$ & [0.1,50] \\
        $kT$ (keV) & $0.231$ & $0.231^{+0.003}_{-0.003}$ & [0.1,1] \\
        Discbb $norm$ & $3.7\cdot10^{5}$ & $3.6^{+0.4}_{-0.3}\cdot10^{5}$ & [1,10$^{8}$] \\
    \end{tabular}
    \label{tab:spectrum_pars}
\end{table}

\subsection{The time-averaged spectrum}

\begin{figure*}
    \centering
    \includegraphics[width=1.025\textwidth]{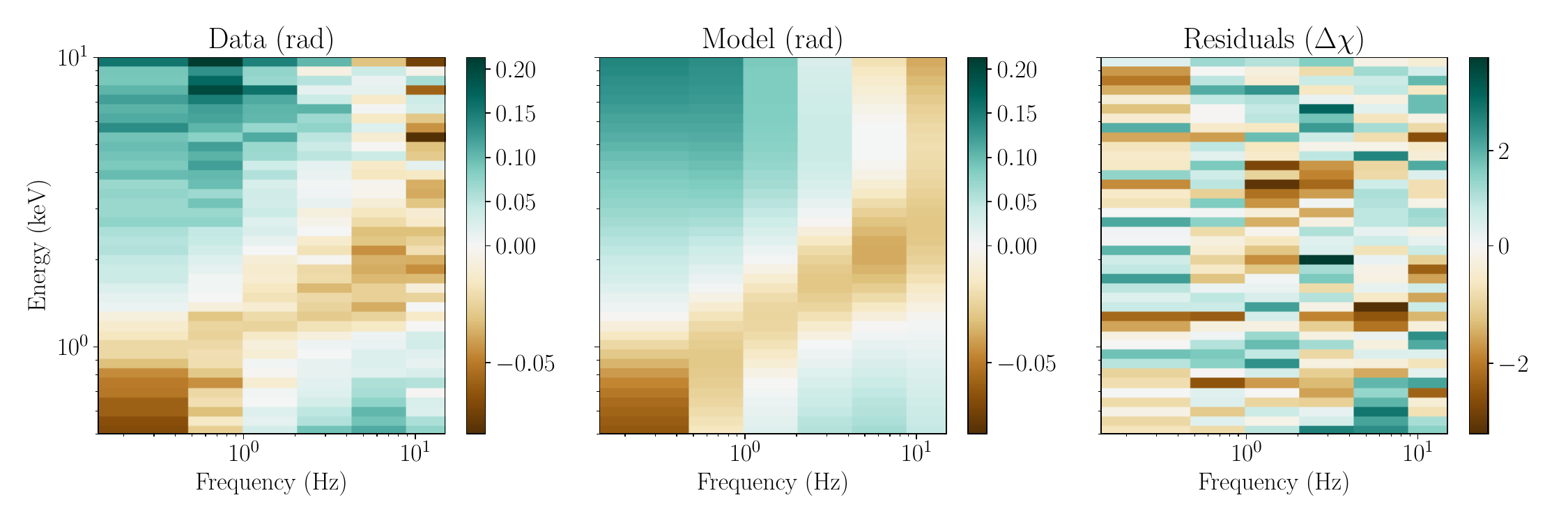}
    \caption{Energy-dependent time lags, showing the data as a function of both the Fourier frequency and energy in units of phase (left plot), the posterior median model in the same units (middle plot), and the residuals in units of $\Delta\chi$ (right plot). Our phenomenological model reproduces the data very well at all energies and Fourier frequencies.}
    \label{fig:lag_fit}
\end{figure*}

\begin{figure}
    \includegraphics[width=\columnwidth]{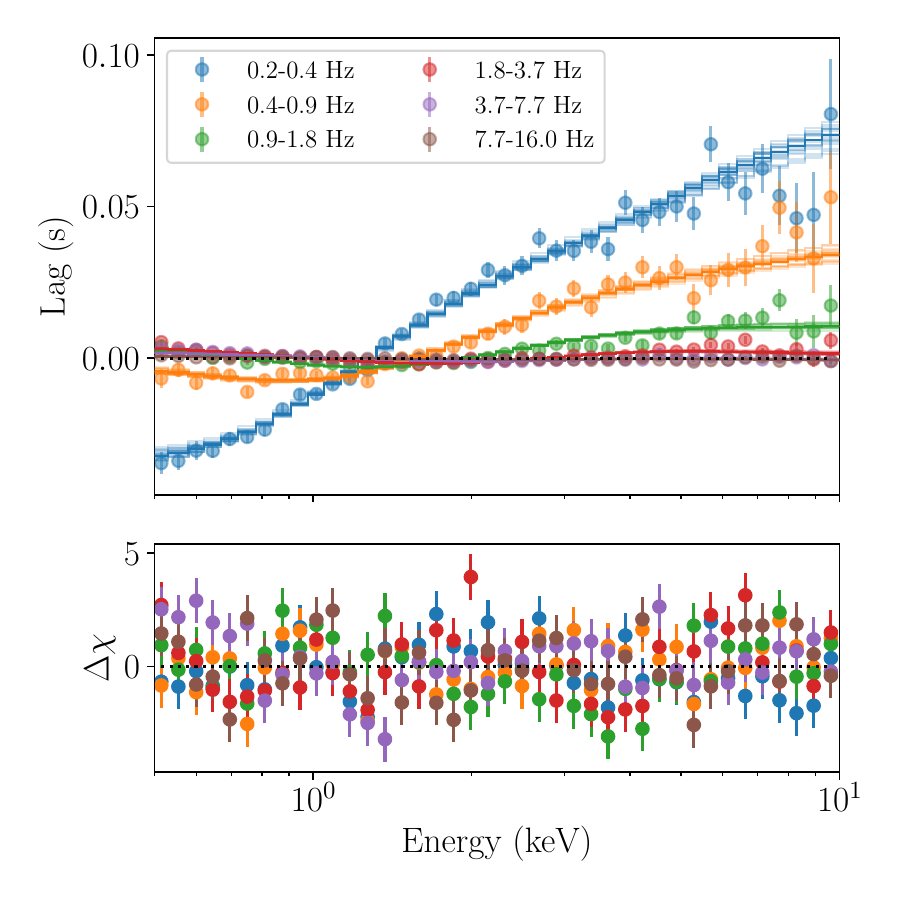}
    \centering
    \caption{Same model shown in fig.\ref{fig:lag_fit} but using a traditional one-dimensional plot to represent the data, posterior median model, and residuals.}
    \label{fig:lag_fit_1d}
\end{figure}

Next, we model the time-averaged spectrum with another common phenomenological model: a standard Shakura-Sunyaev disc \cite{Shakura73} model, combined with a power-law to mimic the Comptonised emission from a corona. Both are modified by the absorption model \textsl{tbabs} \cite{Wilms00}. In the current implementation, we wrap \textsc{PyXspec} to access \textsl{discbb} and \textsl{tbabs}; the functions we use are shown in the software documentation page. As in the case of the power spectrum, we first minimise the $\chi^2$ statistic, and then use the resulting parameter values as an initial guess for running \textsc{emcee}, with the goal of sampling the joint posterior distribution of the parameters. The best-fit statistic of the initial optimisation is $\chi^{2}/\rm{d.o.f.}=572.41/230=2.49$. We use uniform priors for the column density $nH$, photon index $\Gamma$, and disc temperature $kT$, and log-uniform priors for the disc and powerlaw normalisations. We find that the parameter space for this particular case is very simple, and therefore only run the sampler for $6000$ steps using $32$ walkers, with $200$ steps discarded as burn-in. The autocorrelation length for each parameter is $\approx70$ steps, and we thin the chain by a factor $40$ to compute the posteriors. Unlike the case of the power spectrum, we find (unsurprisingly) that our phenomenological model does not represent the data very well. Fig.\ref{fig:spectrum_fit} clearly shows the presence of residuals at soft ($\approx 1$ keV) energies, as well as an emission lines at $\approx 6.4$ keV. These features are likely originating in the reflection component, and modelling them accurately is beyond the scope of this fit, which aims at illustrating a proof of concept for using \textsc{nDspec} and at informing the fits to the cross spectra discussed below. Tab.\ref{tab:spectrum_pars} reports the summary of both the initial minimisation and the 68th percentile credible interval from the posterior sampling. The one- and two-dimensional posterior distributions are shown in Fig.\ref{fig:spectrum_posterior} of the Appendix.

\subsection{Two-dimensional lags and cross spectrum}
\label{sec:crossfit}

Finally, we model the cross spectrum as a function of both energy and Fourier frequency. Because the cross-spectrum is a two-dimensional, complex number this kind of analysis would be particularly cumbersome in existing packages, as users would have to write bespoke code to combine model components and/or account for the instrument response. We begin by fitting phase lags alone, and after finding a good initial model move on to modelling the modulus and phase together. In either case we combine two phenomenological model components (corrected for interstellar absorption with \textsl{tbabs}) described in Sec.\ref{sec:models}: \textsl{pivoting\_pl}, representing a variable corona, plus \textsl{bbody\_bkn}, representing disc irradiation and/or reverberation. Additionally, we include a low energy exponential cut-off in the \textsl{pivoting\_pl} component, mimicking the low energy turnover of a Comptonisation spectrum; this contribution allows us to account for the dilution of the soft lags. The former component is defined explicitly in the Fourier domain as a transfer function, while the latter is defined in the time domain as an impulse response function. As a result, we convert the latter component to a transfer function using the native functionality of a \texttt{CrossSpectrum} class instance. This allows us to directly combine model components by summing them; deriving the appropriate spectral-timing products (e.g. lag-energy spectra) from the total transfer functions is handled automatically by the \texttt{FitCrossSpectrum} instance at each model evaluation. This model can effectively be thought of as the spectral-timing equivalent of a phenomenological powerlaw+discbb spectral fit.

\begin{table}[]
    \centering
    \caption{Inference results for the lag spectra; for the phase correction parameters, the prior column lists the mean and standard deviation of the priors used.}
    \small 
    \begin{tabular}{c|c|c|c}
        Parameter & least-$\chi^{2}$ & Posterior & Prior\\ \hline 
        $\gamma_0$ & $4.8\cdot10^{-2}$ & $4.9^{+0.4}_{-0.4}\cdot10^{-2}$ & [0,0.5]\\
        $s_{\gamma}$ & $7\cdot10^{-3}$ & $-7^{+3}_{-4}\cdot10^{-3}$ & [-0.2,0]\\
        $\phi_0$ & $-1.9$ & $-1.9^{+0.1}_{-0.2}$ & [-3,3]\\
        $s_{\phi}$ & $2.3$ & $2.2^{+0.1}_{-0.2}$ & [-3,3]\\
        $R_0$ & $1.7$ & $1.7^{+0.3}_{-0.2}$ & log-uniform [-3,4] \\
        $kT_0$ & $0.66$ & $0.67^{+0.02}_{-0.02}$ & [0.05,1] \\
        $s_2$ & $1.79$ & $-1.79^{+0.05}_{-0.05}$ & [-10,10] \\
        $s_t$ & $-0.41$ & $-0.41^{+0.01}_{-0.01}$ & [-3,0] \\ \hline        
        $\Phi_1$ & $-1\cdot10^{-3}$ & $-1^{+1}_{-1}\cdot10^{-3}$ & [$\mu=0$, $\sigma=0.05$] \\
        $\Phi_2$ & $9\cdot10^{-3}$ & $9^{+2}_{-2}\cdot10^{-3}$ & [$\mu=0$, $\sigma=0.05$] \\
        $\Phi_3$ & $1.4\cdot10^{-2}$ & $1.5^{+0.1}_{-0.1}\cdot10^{-2}$ & [$\mu=0$, $\sigma=0.05$] \\
        $\Phi_4$ & $1.8\cdot10^{-2}$ & $1.8^{+1}_{-1}\cdot10^{-2}$ & [$\mu=0$, $\sigma=0.05$] \\
        $\Phi_5$ & $1.5\cdot10^{-2}$ & $1.6^{+2}_{-2}\cdot10^{-2}$ & [$\mu=0$, $\sigma=0.05$] \\
        $\Phi_6$ & $7\cdot10^{-3}$ & $7^{+2}_{-2}\cdot10^{-3}$ & [$\mu=0$, $\sigma=0.05$] \\
    \end{tabular}

    \label{tab:lags_pars}
\end{table}

To summarise, the parameters of the model we use here are:
\begin{description}
\item [$n_H$] absorption column density 
\item [$\Gamma$] power-law photon index 
\item [$A_0$] power-law normalisation
\item [$\gamma_0$] fractional amplitude of the variability in $\Gamma$ at $\nu_0$=0.2 Hz 
\item [$s_{\gamma}$] frequency scaling parameter for the fractional amplitude of the variability in $\Gamma$ 
\item [$\phi_0$] initial phase between changes in the power-law normalisation and photon index at $\nu_0=0.2$ Hz 
\item [$s_{\phi}$] frequency scaling parameter for the phase between the variability in $\Gamma$ and $A$ 
\item [$E_{cut}$] low energy cutoff in the powerlaw emission
\item [$R_0$] disc irradiation IRF normalisation 
\item [$kT_0$] disc IRF initial temperature
\item [$t_{brk}$] disc IRF rise timescale 
\item [$s_1$] disc IRF rise slope 
\item [$s_2$] disc IRF decay slope 
\item [$s_t$] disc temperature time scaling slope   
\end{description}

We freeze both $n_H$ and $\Gamma$ to the values we found when fitting the time-averaged spectrum, as the latter constrains these quantities significantly more strongly than the cross spectrum. We also tie the disc IRF temperature to the low energy cut-off $E_{cut}$ in the power-law spectrum. When modelling phase lags we fix the power-law normalisation to unity; this is because the phase of the cross spectrum is sensitive to the relative importance of the pivoting power-law component with respect to the reverberation signal, but not to their absolute values. In exploratory models we found that $s_1$ and $t_{brk}$ are un-constrained by the data; this is not surprising given that they mostly impact shorter timescales and higher Fourier frequencies than the 16 Hz our analysis is limited to. As a result, we fixed $t_{brk}=0.01$ s (a third of the time resolution of the lightcurves we extracted) and $s_1=4$. Additionally, we found that enabling the phase re-normalisation discussed in Sec.\ref{sec:optimization} improved the quality of all fits without affecting the other parameters, so in the paper we only report fits in which it has been enabled.

We begin as previously with least-squares optimisation, without enabling the phase re-normalisation discussed in Sec.\ref{sec:optimization}. The results are shown in Fig.\ref{fig:lag_fit} and \ref{fig:lag_fit_1d}; we include both one- and two-dimensional plots to highlight the visualisation features built into \textsc{nDspec}. The best-fitting statistic is $\chi^{2}/\rm{d.o.f.}=397.92/226=1.76$.  We use uniform priors for all the model parameters, except the reverberation normalisation $R_0$, for which we use a log-uniform prior, and the phase correction factors. We expect the latter to be small (and ideally zero, corresponding to perfect instrument calibration), so we use Gaussian priors centred around $0$, with a standard deviation of $0.05$ radians. As before, we initialise an \textsc{emcee} sampler from this minimum and run the sampler until well after convergence using 32 walkers; this time, the integrated auto-correlation time is $\approx200$ steps, and therefore we evolve the chain for 15000 steps. We then calculate the posterior distributions by thinning the chain by a factor 150.. As is the case for the time-averaged spectrum, we find that the parameter space is very well behaved, with all the parameters well constrained. The best-fit parameters and posterior median and 68th percentile credible intervals are reported in Tab.\ref{tab:lags_pars}, and the posterior distribution of all the model parameters is shown in Fig.\ref{fig:lags_posterior} of the Appendix.

\begin{table}[]
\centering
\caption{Inference results for the full cross spectra; for the phase correction parameters, the prior column lists the mean and standard deviation of the priors used.}
\small
    \begin{center}
        \begin{tabular}{c|c|c|c}
        Parameter & least-$\chi^{2}$ & Posterior & Prior \\ \hline 
        $A_0$ & $6.2$ & $6.7^{+0.1}_{-0.1}$ & log-uniform [-3,4]\\
        $\gamma_0$ & $8.5\cdot10^{-2}$ & $8.5^{+0.3}_{-0.3}\cdot10^{-2}$ & [0,0.5] \\
        $s_{\gamma}$ & $-1.7\cdot10^{-2}$ & $-1.70^{+0.03}_{-0.03}\cdot10^{-2}$ & [-0.2,0] \\
        $\phi_0$ & $-1.9$ & $-1.95^{+0.07}_{-0.06}$ & [-3,3] \\
        $s_{\phi}$ & $2.3$ & $2.25^{+0.08}_{-0.07}$ & [-3,3] \\
        $R_0$ & $96$ & $115^{+11}_{-9}$ & log-uniform [-3,4] \\
        $kT_0$ & $0.272$ & $0.272^{+0.003}_{-0.004}$ & [0.05,1] \\
        $s_2$ & $-2.1$ & $-2.10^{+0.05}_{-0.05}$ & [-10,10] \\
        $s_t$ & $0$ & $-0^{+*;*}_{-0.001}$ & [-3,0] \\ \hline        
        $\Phi_1$ & $-1.1\cdot10^{-2}$ & $-1.1^{+0.1}_{-0.1}\cdot10^{-2}$ & [$\mu=0$, $\sigma=0.05$] \\
        $\Phi_2$ & $8\cdot10^{-3}$ & $7^{+1}_{-2}\cdot10^{-3}$ & [$\mu=0$, $\sigma=0.05$] \\
        $\Phi_3$ & $1\cdot10^{-3}$ & $0^{+1}_{-1}\cdot10^{-3}$ & [$\mu=0$, $\sigma=0.05$]  \\
        $\Phi_4$ & $1.1\cdot10^{-2}$ & $1.1^{+0.2}_{-0.2}\cdot10^{-2}$ & [$\mu=0$, $\sigma=0.05$] \\
        $\Phi_5$ & $1.6\cdot10^{-2}$ & $1.6^{+0.2}_{-0.2}\cdot10^{-2}$ & [$\mu=0$, $\sigma=0.05$] \\
        $\Phi_6$ & $1.4\cdot10^{-2}$ & $1.4^{+0.2}_{-0.2}\cdot10^{-2}$ & [$\mu=0$, $\sigma=0.05$] \\ \hline 
        $m_1$ & $1.15$ & $0.80^{+0.05}_{-0.05}$ & [$\mu=1$, $\sigma=0.15$]  \\
        $m_2$ & $1.90$ & $1.38^{+0.09}_{-0.08}$ & [$\mu=1$, $\sigma=0.15$] \\
        $m_3$ & $1.41$ & $0.99^{+0.06}_{-0.06}$ & [$\mu=1$, $\sigma=0.15$] \\
        $m_4$ & $1.04$ & $0.73^{+0.05}_{-0.04}$ & [$\mu=1$, $\sigma=0.15$] \\
        $m_5$ & $1.20$ & $0.84^{+0.05}_{-0.05}$ & [$\mu=1$, $\sigma=0.15$] \\
        $m_6$ & $1.56$ & $1.10^{+0.07}_{-0.07}$ & [$\mu=1$, $\sigma=0.15$] \\  
        \end{tabular}
    \end{center}    
    *: parameter pinned to its limit
    \label{tab:cross_pars}
\end{table}

\begin{figure*}
    \centering
    \includegraphics*[width=0.85\textwidth]{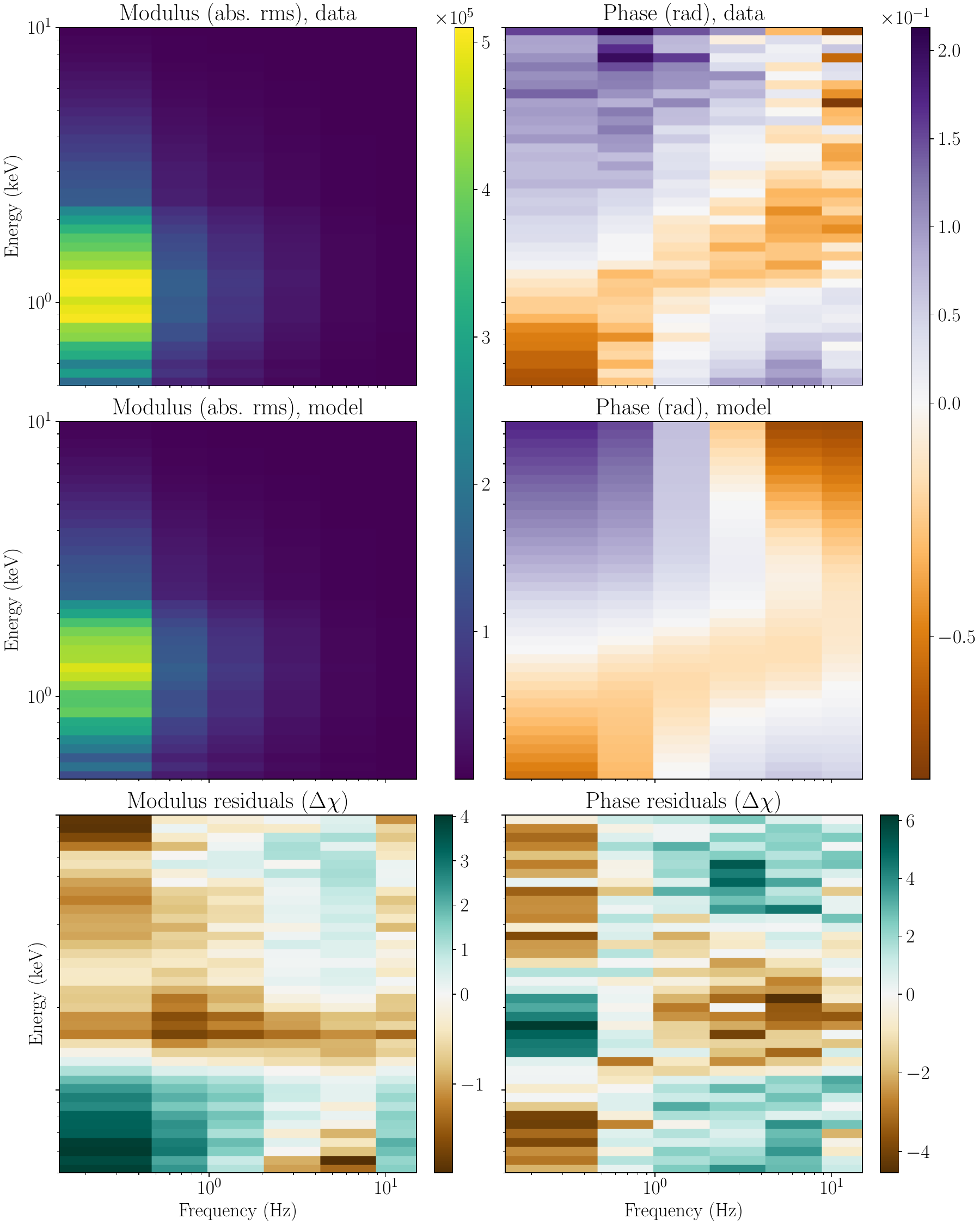}
    \caption{Full cross spectrum, shown as a two-dimensional plot. The top row is the data as a function of Fourier frequency and energy, the middle row shows the posterior median model in the same units, and the bottom row shows the residuals in units of $\Delta\chi$. The left column shows the modulus (in units of absolute rms) and the right column shows the phase (in units of radians). The model shows strong structured residuals in both modulus and phase, and therefore it fails to reproduce the data.}
    \label{fig:cross_fit}
\end{figure*}

\begin{figure*}
    \centering
    \includegraphics[width=\textwidth]{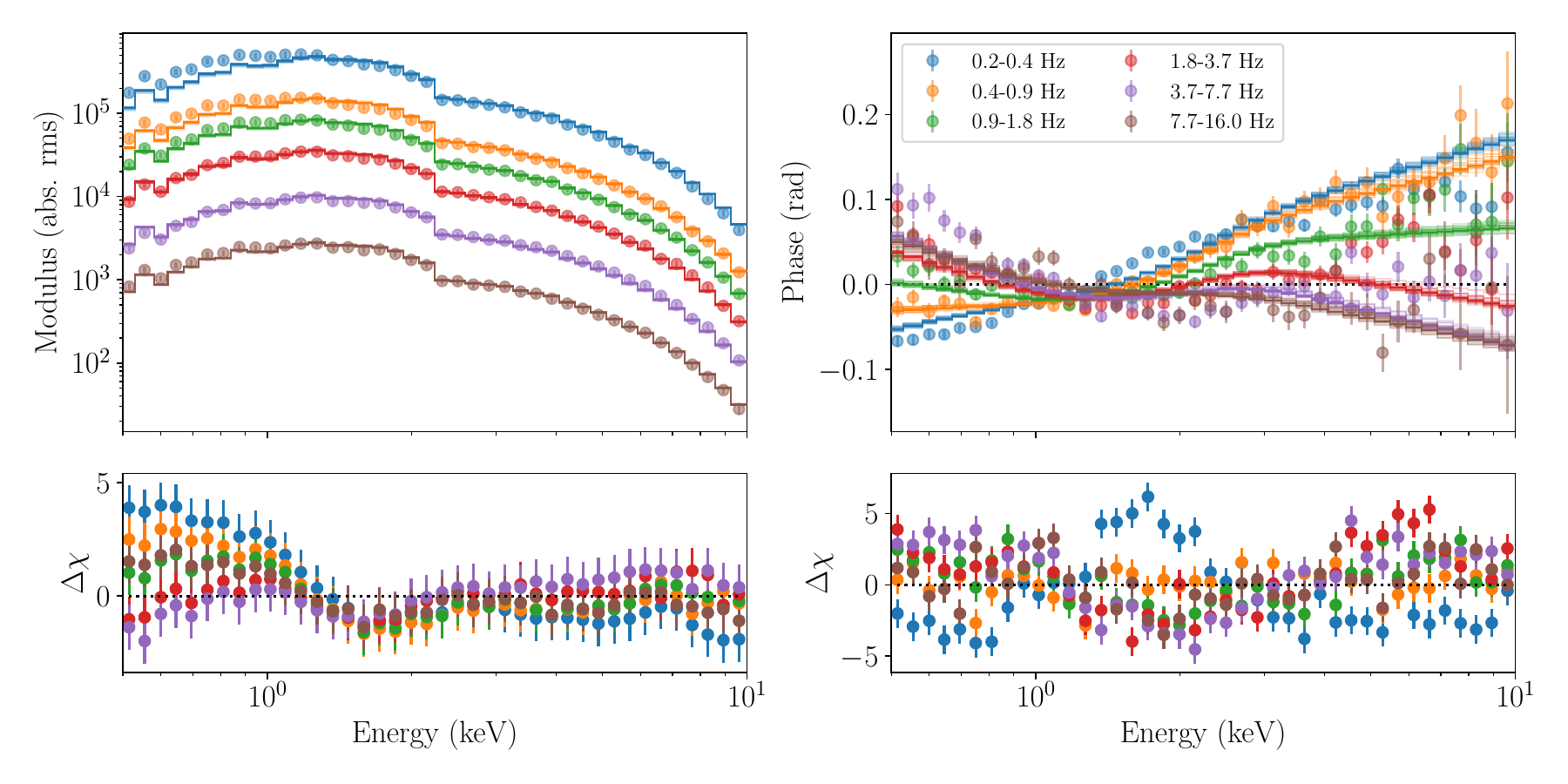}
    \caption{Same model shown in fig.\ref{fig:cross_fit}, but using a traditional one-dimensional plot to represent the data, posterior median model, and residuals. The left column shows the modulus in units of absolute rms, and the right column shows the phase in units of radians.}
    \label{fig:cross_fit_1d}
\end{figure*}

Finally, we attempt to fit the energy and frequency dependent modulus and phase of the cross spectrum. However, in this case our phenomenological model is not sufficient to reproduce the data in its complexity. 

We initialise our fit from the best-fitting parameters for the phase lags, and free the pivoting power-law normalisation $A_0$. We also enable re-normalising both the phase and modulus of the cross spectrum (Sec.\ref{sec:optimization}). In preliminary fits, we found that due to the much higher S/N ratio of the modulus compared to the phase, we had to include a $0.75\%$ systematic error on the former; otherwise, the optimiser would effectively ignore the phase. 

We utilise the same priors as with the previous model; additionally, we set a log-uniform prior for the pivoting power-law normalisation $A_0$, and Gaussian priors for the modulus re-normalisation factors. We expect the latter to be as close to unity as possible (meaning no need for re-normalising the modulus), so we use Gaussians centred around unity with a standard deviation of $0.15$. 

As for previous models, we sample the posterior distribution of our model parameters after finding a minimum of the fit statistic.  The best-fitting statistic found by the optimiser is  $\chi^{2}/\rm{d.o.f.}=1373.30/459=2.99$, indicating a poor fit. The increased number of free parameters ($A_0$ and the frequency-dependent re-normalisation constants, $m_1$ through $m_6$) increases the complexity of the parameter space, and as a result we run a sampler with more walkers ($48$) and for longer ($25000$ steps). The autocorrelation length after burn-in is $\approx500$ steps. 

Results are shown in Fig.\ref{fig:cross_fit} and \ref{fig:cross_fit_1d}, again showing two- and one-dimensional plots respectively. The model currently reproduces the phase lags only qualitatively, with energy-dependent residuals appearing; more importantly, it completely fails at capturing the energy dependence of the modulus in all frequency bins. In particular, the residuals systematically show an excess at soft X-ray energies and one around $\approx 6$ keV, similarly to the reflection signature in a time-averaged spectrum. Similar features also appear less clearly in the phase residuals. These are likely due to relativistic reverberation (possibly due to light travel times) between the coronal continuum and the reflection component, which our model does not account for.

The inferred parameters from both optimising the fit statistic and sampling the posterior distribution are in Tab.\ref{tab:cross_pars}. In general, we both optimisation and posterior inference agree well, with the exception of the normalisations $A_0$ and $R_0$ as well as the modulus normalisation constants. This is caused by these parameters being degenerate with each other, causing the initial optimiser to be stuck in a local minimum. Finally, we note that in this section we model the modulus and phase for ease of visualisation and interpretation, but it is also possible for users to model the real and imaginary parts instead, with identical results. 

In conclusion, we show that \textsc{nDspec} enables straightforward modelling of spectral-timing products with ease, especially compared to approaches implemented in existing packages. Because \textsc{nDspec} directly integrates with the \textsc{emcee} package rather than reimplement it, we also have direct access to additional functionality in that package not available existing software such as \textsc{XSPEC}, like the autocorrelation time used to diagnose MCMC convergence or the differential evolution moves. The phenomenological models included in \textsc{nDspec} can roughly reproduce many of standard observational spectral-timing products, such as power spectra, time-averaged spectra, and energy-dependent lags in multiple Fourier frequency bins. However, they are not sufficient to capture the full extent of the information encoded in the cross spectrum of a NICER observation. Fully modelling the energy dependence of the variability amplitude likely requires a relativistic reverberation component and more realistic treatment of the accretion variability, the implementation of which is beyond the scope of the initial release of the software. 

\section{Future development and conclusions}
\label{sec:development} 

In this paper, we have detailed the the initial version of a new Python-based X-ray modelling library, which we called \textsc{nDspec}. It provides a framework for high-dimensional X-ray data and interface layers to state-of-the-art packages designed for modern statistical inference. The current version of the software allows users to model individual instances of three data products common to X-ray astronomy -- power spectra, cross spectra, and limited support for time-averaged spectra -- and to perform Bayesian sampling of model parameter space by directly linking model evaluations and parameters with the \textsc{emcee} library. The software also provides utilities to fold and unfold mono- and multi-dimensional models through instrument response matrices, as well as a class to carry out spectral-timing model evaluations. While in its current form, by necessity, it duplicates some functionality implemented in part in other, existing packages, it will only really reveal its power once going beyond this initial set of functionality. Going forward, we are planning several main development paths for \textsc{nDspec}.

In the near future, we plan to expand the statistical back end of the software. In particular, we plan on supporting non-Gaussian likelihoods, such as the Whittle, Poisson, or Nathan or modified Laplace distributions, all of which are appropriate to specific spectral-timing datasets \citep{Vaughan10, Cash79, Nathan26}. We also plan on supporting nested sampling (\citealt{Multinest1}, \citealt{Ultranest}, \citealt{Dynesty}) and  ensemble slice sampling \citep{zeus} as complements to the \textsc{emcee} sampler we used for this paper, and to take account of the fact that modern X-ray data contain idiosyncrasies (such as multi-modal, degenerate, or highly dimensional parameter spaces) not well addressed by a single standard sampling algorithm. In addition, we plan to support jointly modelling multiple datasets (e.g. from multiple instruments), and enable users to use Xspec models natively in Python while bypassing PyXspec completely, which provides a considerable performance improvement. With the exception of the Whittle and modified Laplace likelihood, all of these features are complete and in testing at the time of writing; we leave their discussion for a future paper that is currently in preparation.

Beyond these initial improvements, we plan on extending the types of data that can be modelled. A future version will include support for polarimetric data (including both spectro-polarimetry and polarimetric-timing analysis, e.g. \citealt{Ixpe1}, \citealt{Ingram17}, \citealt{Ewing25}), and extend the types of spectral-timing data that can be modelled (e.g. phase-resolved energy spectra and dynamical power spectra). In practice, this means implementing fitter objects to handle these different types of data. We aim to implement polarimetry and generic two-dimensional fitters within one year. In addition, we plan on explicitly supporting multi-wavelength data. 

Third, we plan to further optimise the performance of \textsc{nDspec} by offloading part of the computational load to a GPU. In a Python framework, this is relatively easily implemented with a library such as \textsc{Jax} \citep{jax}. We are going to work with a software support team in the following months, with the goal of profiling the overall code and implementing optimisations such as GPU acceleration where appropriate (for instance, when handling large response matrices, as is the case of micro-calorimeters). In the longer term, an added benefit of utilising a library such as \textsc{Jax} is the ability to perform auto-differentiation. Auto-differentiable code provides the gradient of the function being computed in addition to the function itself, and it is necessary for implementing machine-learning methods or Hamilton Monte Carlo samplers (\citealt{Girolami11}, \citealt{Cranmer20}, \citealt{Buchner23b}). We acknowledge that re-building the Xspec model library to be auto-differentiable is a large endeavour (though this task has already begun to be undertaken in \texttt{Jaxspec} \citet{jaxspec}); on the other hand, timing and polarimetry models are both more limited in their availability and more computationally expensive than spectral models. Our hopes are that: a) by providing a modern software tool tailored to modelling these complex datasets, the community will build a model library as was the case for Xspec and X-ray spectroscopy; and b) these models will be designed from the start to take advantage of modern computational techniques, such as auto-differentiability. As an open source software, our long-term vision for continued support and development is identical to those of other open source packages such as \textsc{Stingray} and \textsc{Astropy}: rely on community contributions for specific features, while maintaining a core team of developers to organise development, version control, and releases, and additional capabilities.

\section{Data Availability}
The code is entirely open source under an MIT licence and fully documented, and can be found at \href{https://github.com/nDspec/nDspec}{https://github.com/nDspec/nDspec}. A reproduction package to re-create every plot and re-run every fit in the paper can be found at DOI: \href{https://zenodo.org/records/16567436}{10.5281/zenodo.16567436}.

\begin{acknowledgements}
\textsc{nDspec} makes use of \textsc{Astropy} (\citealt{astropy13}, \citealt{astropy18}, \citealt{astropy13}), \textsc{NumPy} \citep{numpy}, \textsc{SciPy} \citep{scipy}, \textsc{Stingray} (\citealt{Stingray1}, \citealt{Stingray1}), \textsc{LMFIT} \citep{lmfit}, and \textsc{matplotlib} \citep{matplotlib}. 
\end{acknowledgements}

\bibliography{references}
\bibliographystyle{aa}


\begin{appendix} 
\section{The model library}
\label{sec:models}

The last feature included in the initial release of nDspec is a small library of timing and/or spectral-timing phenomenological models. As is the case for existing \textsc{Xspec} models, each model can be used as a separate component and combined as necessary to fit a given dataset. In this section, we describe all the models currently included in the library. The notation used is as follows. We use $E$ to denote the independent variable being photon energy, $\nu$ for Fourier frequency, $t$ for time, and $x$ for when it is arbitrary. We use $f$ to denote models that return arbitrary units, $n$ for models which return a photon flux, $g$ for models of impulse response functions, and $G$ for models of transfer functions. 

The one-dimensional models included in the initial release are:

\noindent \textbf{lorentz}: this is a standard Lorentzian, defined as
\begin{equation}
    f_{\rm lor}(x) = \frac{2R^{2}Qx_{res}}{\pi\left[x^{2}_{res}+4Q(x-x_{res})^{2}\right]},
    \label{eq:lorentz}
\end{equation}
where $x_{res}=x_{\rm peak}\sqrt{1+1/4Q{^2}}$ is the resonance scale given a peak at $x_{peak}$ and a quality factor $Q$, and (when the model is used in the Fourier domain) $R=rms/\sqrt{0.5-\tan^{-1}(-2Q)/\pi}$ is the normalisation for a given fractional rms $rms$. In nDspec, we use $x_{peak}$, $Q$ and $rms$ as the free parameters, as they are more immediately interpretable than $R$ or $x_{res}$ when fitting power spectra.

\noindent\textbf{cross\_lorentz}: this is a Lorentzian modified by an arbitrary phase, defined as
\begin{equation}
    f_{\rm clor}(x) = f_{\rm lor}(x)e^{i\phi},
    \label{eq:cross_lorentz}
\end{equation}
where $f_{\rm lor}(x)$ is identical to the Lorentzian function defined in Eq.\ref{eq:lorentz} and $\phi$ is an additional free parameter used to include a phase in the Lorentzian. This model component can be thought of as an extension of the previous one, when one is interested in fitting data encoded in complex rather than real numbers \citep{Mendez24}.

\noindent\textbf{powerlaw}: a standard powerlaw, defined as
\begin{equation}
    f_{\rm pl}(x) = Ax^{\Gamma},
    \label{eq:powerlaw}
\end{equation}
where $A$ is the normalisation and $\Gamma$ is the slope. We note that due to the generality of the model, we omit the minus sign before the slope typically used in X-ray spectral models.

\noindent\textbf{brokenpower}: a smoothly broken power-law, defined as
\begin{equation}
    f_{\rm bp}(x) = A\frac{\left(x/x_{brk}\right)^{\Gamma_1}}{1+\left(x/x_{brk}\right)^{\Gamma_1-\Gamma_2}},
    \label{eq:bknpower}
\end{equation}
where $A$ is the normalisation and $\Gamma_1$ and $\Gamma_2$ are the slopes before and after $x_{brk}$ respectively. 

\noindent\textbf{gaussian}: a standard Gaussian profile, defined as
\begin{equation}
    f_{\rm gauss}(x) = \frac{A}{\sigma\sqrt{2\pi}}\exp{\left[-\left(x-x_c\right)^2/2\sigma^2\right]}
    \label{eq:gauss},
\end{equation}
where $A$ is the normalisation of the Gaussian, $\sigma$ its width, and $x_c$ its centre.

\noindent\textbf{bbody}: a black body, defined to have identical normalisation to the \textsc{Xspec} implementation:
\begin{equation}
    n_{\rm bb}(E) = 8.0525\frac{AE^2}{(k_bT)^4\left(\exp{E/k_bT-1}\right)},
    \label{eq:bbody}
\end{equation}
where $A$ is the normalisation, $E$ is the photon energy, $k_b$ the Boltzmann constant, and $T$ the temperature in expressed as $k_bT$, in units of keV.

\noindent\textbf{varbbody}: a model accounting for the linear changes in a black body spectrum in response to a fluctuation in temperature \citep{vanParadijs86,Uttley25}:
\begin{equation}
    n_{\rm vbb}(E) = 2.013\frac{AE^3\exp{E/k_bT}}{(k_bT)^5\left(\exp{E/k_bT-1}\right)^2},
    \label{eq:varbbody}
\end{equation}
where the free parameters are identical to eq.\ref{eq:bbody}. This model is expected to be used as part of a model for a linear impulse response function. 

Each of the components above is provided as a Python function, users are free to combine them as necessary to form more complex and/or multi-dimensional models. We also include several two-dimensional models, which are intended (but not limited) to be used as phenomenological impulse response or transfer functions. These are:

\noindent\textbf{gauss\_fred}: a phenomenological impulse response function, in which the energy dependence is specified by a Gaussian profile (Eq.\ref{eq:gauss}) which narrows over time, and the normalisation changes as a fast rise, exponential decay profile. The IRF is defined as
\begin{equation}
    g_{\rm gaussf}(E,t) = A(t)\frac{\exp{\left[-\left(E-E_c\right)^2/2\sigma(t)^2\right]}}{\sigma(t)\sqrt{2\pi}}
    \label{eq:fred_gauss}
\end{equation}
the normalisation $A(t)$ changes over time as:
\begin{equation}
    A(t) = A_0 \exp(-t_r/t)\exp(t/t_d),
\end{equation}
where $A_0$ is the initial normalisation, $t_r$ the rise timescale, and $t_d$ the decay timescale. The width of the line $\sigma(t)$ follows a power-law in time, normalised to the initial width $\sigma_0$:
\begin{equation}
 \sigma(t) = f_{\rm pl}(t/t_0),  
\end{equation}
where $f_{\rm pl}(t/t_0)$ is given by Eq.\ref{eq:powerlaw}, using an initial width $\sigma_0$ to set the normalisation and a slope $s$ to set the variation of the line width over time. $t_0$ is the first time bin of the grid over which the model is computed. By taking the initial line centre $E_c\approx 6.5$ keV, one might use this IRF to roughly replicate the behaviour of a relativistic iron line profile. 

\noindent\textbf{bbody\_bkn}: a phenomenological model similar to that described above. However, in this case the time dependence of the normalisation is a broken power-law (Eq.\ref{eq:bknpower}), and the energy dependence is due to a varying black body (Eq.\ref{eq:varbbody}) whose temperature cools over time:
\begin{equation}
    g_{\rm bb bp} (E,t) = f_{\rm bp}(t/t_0) n_{\rm vbb} (E),
\end{equation}
Here, the time-dependent normalisation $f_{\rm bp}(t/t_0)$ follows a smoothly broken powerlaw slope as defined in eq.\ref{eq:bknpower}; its input parameters are the overall normalisation $A_0$, the break timescale $t_{brk}$, and the rise and decay slopes $s_1$ and $s_2$, defined before and after $t_{brk}$ respectively. The energy dependence of the IRF is a variable black body as defined in eq.\ref{eq:varbbody}, where the temperature is additionally assumed to change over time following a powerlaw \ref{eq:powerlaw}:
\begin{equation}
    T(t/t_0) = f_{\rm pl}(t/t_0);
\end{equation}
we use $T_0$ to denote the initial temperature and $s_t$ to denote the slope of this power-law. Similarly to the model above, this model is meant to mimic a reverberation signature, for example due to disc irradiation. 

\noindent\textbf{gauss\_bkn} and \textsl{bbody\_fred}: these models are identical to the two above, except the functions controlling the energy and/or time dependence are inverted (e.g. a fast rise, exponential decay for a black body energy dependence, and vice versa). 

\noindent\textbf{pivoting\_pl}: A model of a pivoting power-law, similar to that of the Reltrans model \citep{Mastroserio18,Mastroserio21} and designed to represent phenomenologically the variability signatures caused by a e.g. a corona responding to variable heating and cooling in an accretion flow \cite{Uttley25}. In this model, the energy spectrum is described by a power-law; both the normalisation and the photon index vary over time, leading to energy-dependent correlated variability and phase lags. In particular, the variations in the photon index are small enough compared to the normalisation that the energy-dependent transfer function can be derived from a first-order Taylor expansion of the coronal spectrum. Under these assumptions, the transfer function is given by
\begin{equation}
    G(E,\nu) = f_{\rm pl}(E)\left[1-\gamma(\nu)e^{i\Phi_{AB}(\nu)}\right],
\end{equation}
where the energy dependent term $f_{\rm pl}(E)$ is a power-law of normalisation $A$ and slope $\Gamma$ (Eq.\ref{eq:powerlaw}), $\gamma(\nu)$ is the amplitude of fluctuations in $\Gamma$ as a fraction of the amplitude of the fluctuations in $A$, and $\phi_{AB}(\nu)$ is the phase between changes in the power-law index and normalisation. Following \cite{Mastroserio18,Mastroserio21} $\phi_{AB}(\nu)>0$ produces hard lags, and vice versa $\phi_{AB}(\nu)<0$ produces soft lags. For a more quantitative discussion on interpreting $\gamma(\nu)$ and $\phi_{AB}(\nu)$, see the discussion in Sec.3.1 in \citealt{Mastroserio21}. The only difference between the \textsc{Reltrans} and \textsc{nDspec} implementations is in the frequency dependence of $\gamma(\nu)$ and $\phi_{AB}(\nu)$. In \textsc{Reltrans} these terms are left to vary freely in each Fourier frequency bin. In \textsc{nDspec} we introduce a more restrictive prescription on both terms, in order to limit the number of free parameters. We define:
\begin{align}
    \gamma(\nu) = \gamma_0 + \log_{10}(\nu/\nu_0)^{s_{\gamma}} \\
    \phi_{AB}(\nu) = \phi_{AB,0} + \log_{10}(\nu/\nu_0)^{s_{\phi}};
    \label{eq:scaling_nu}
\end{align}
meaning that both $\gamma(\nu)$ and $\phi_{AB}(\nu)$ vary smoothly as a function of the logarithm of Fourier frequency, with respect to some initial frequency $\nu_0$ (here, we use the lowest frequency bin in the cross spectrum). $\gamma_0$ and $\phi_{AB,0}$ set the values at that initial frequency bin, while $s_{\gamma}$ and $s_{\phi}$ set the dependence on the frequency. 

\section{Posterior distributions}
\label{sec:appendix}
\begin{figure*}
\centering
\includegraphics[width=\textwidth]{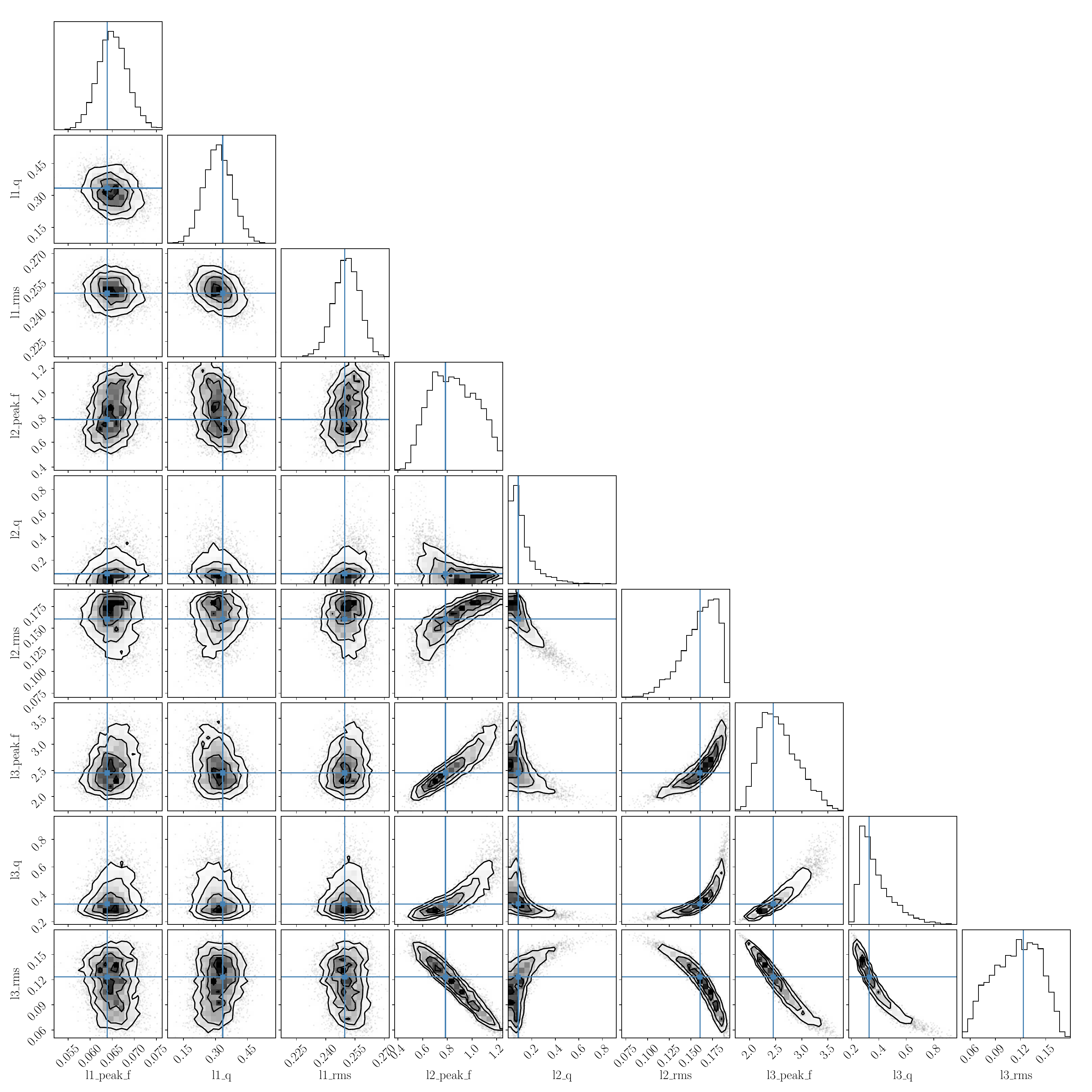}
\caption{Posterior distribution for the power spectrum. Blue lines show the values found by the least chi square optimisation. }
\label{fig:psd_posterior}
\end{figure*}

\begin{figure*}
\centering
\includegraphics[width=\textwidth]{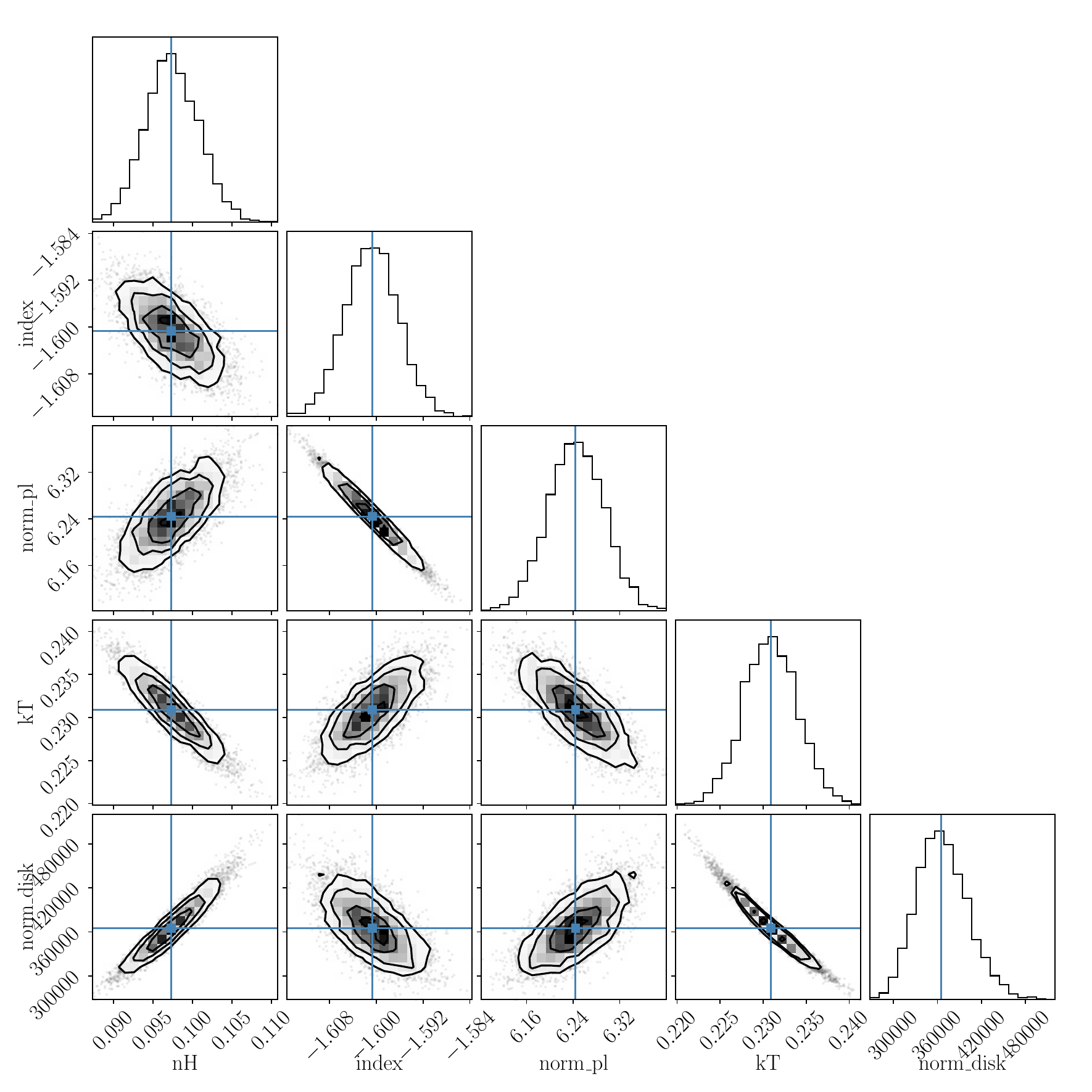}
\caption{Posterior distribution for the time-averaged spectrum. Blue lines show the values found by the least chi square optimisation. }
\label{fig:spectrum_posterior}
\end{figure*}

\begin{figure*}
\centering
\includegraphics[width=\textwidth]{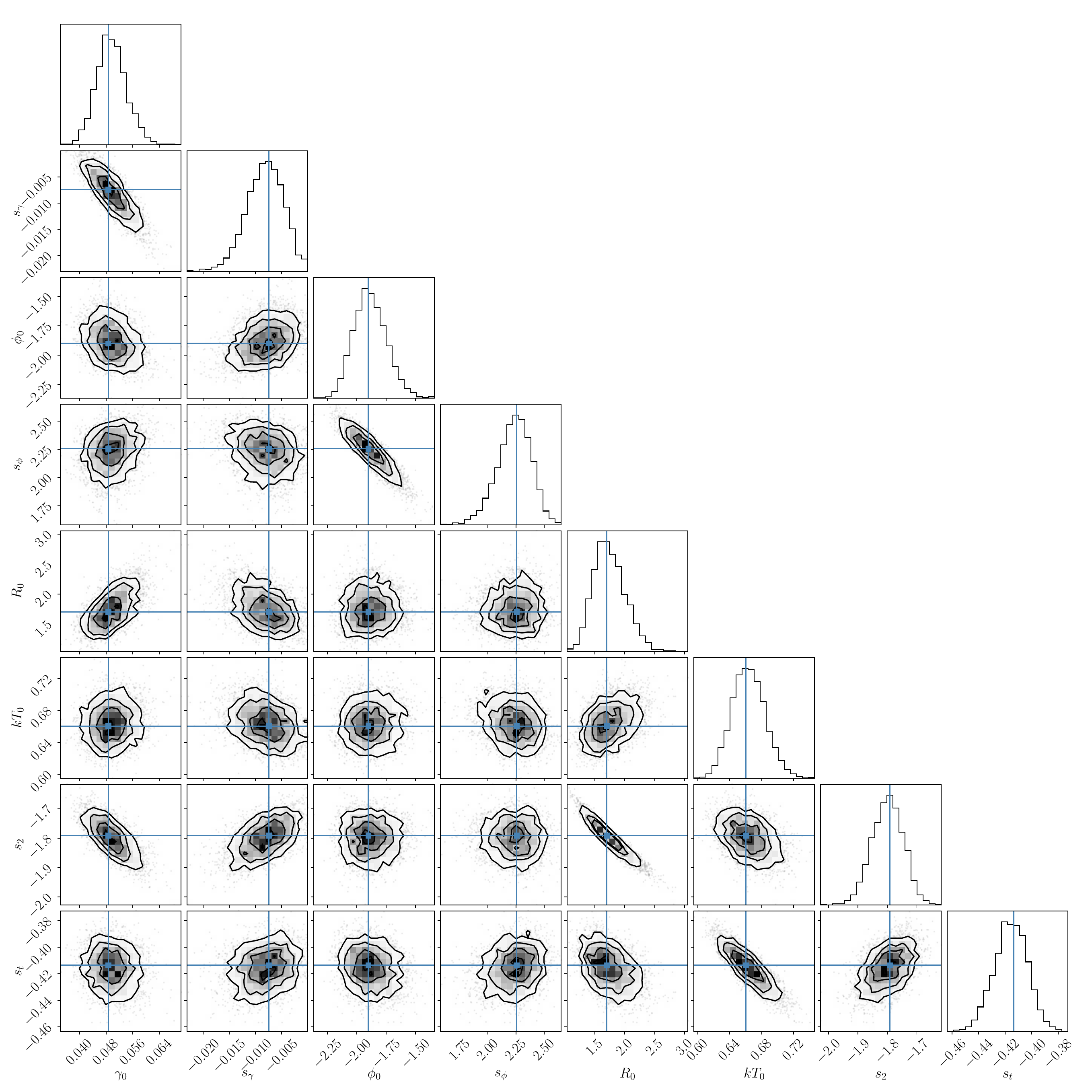}
\caption{Posterior distribution for the energy-dependent lag spectra as a function of Fourier frequency. Blue lines show the values found by the least chi square optimisation. The phase re-normalisation constants are not shown for clarity, but they can be found on the accompanying Zenodo repository.}
\label{fig:lags_posterior}
\end{figure*}

\begin{figure*}
\centering
\includegraphics[width=\textwidth]{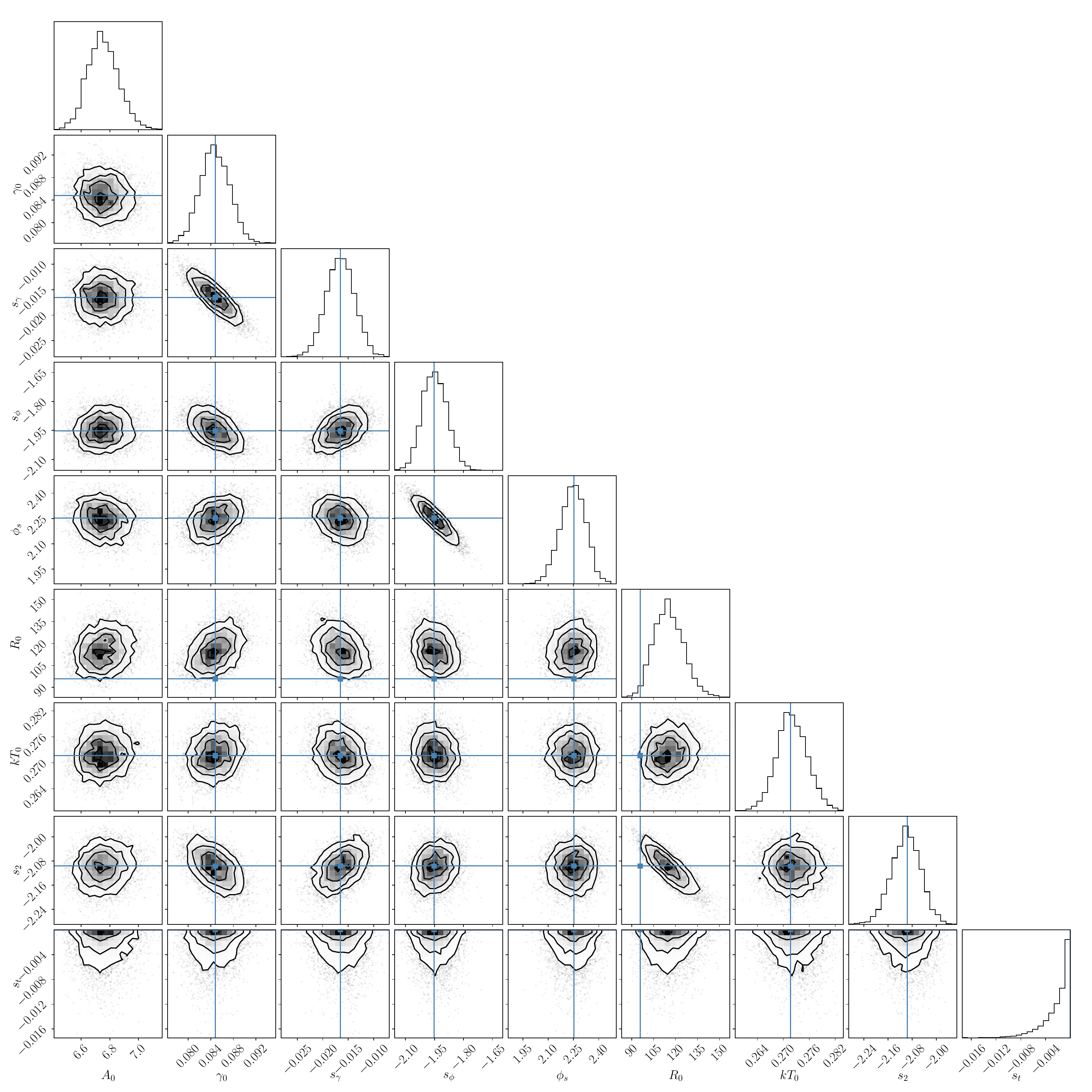}
\caption{Posterior distribution for the fit of the energy-dependent cross spectra as a function of Fourier frequency. Blue lines show the values found by the least chi square optimisation. Both the phase and modulus re-normalisation constants are not shown for clarity, but they can be found on the accompanying Zenodo repository.}
\label{fig:cross_posterior}
\end{figure*}

Here we report the posterior distributions for all the models. Fig.\ref{fig:psd_posterior} shows the corner plot for the fit of the power spectrum. Fig.\ref{fig:spectrum_posterior} shows the corner plot for the fit of the time-averaged spectrum. Fig.\ref{fig:lags_posterior} shows the corner plot for the fit of the lag-energy spectra. Fig.\ref{fig:cross_posterior} shows the corner plot for the fit of the cross spectrum. 

\end{appendix}

\label{LastPage}
\end{document}